\documentclass[journal]{IEEEtran}
\ifCLASSINFOpdf
\else
\fi
\usepackage{amsmath,amssymb}
\usepackage{optidef}
\usepackage{tikz}
\usepackage{pgfplots}
\usetikzlibrary{calc}
\usetikzlibrary{intersections}
\usetikzlibrary{shapes.geometric}
\usetikzlibrary{arrows}
\usetikzlibrary{arrows.meta,arrows}
\usetikzlibrary{patterns} 
\newcommand\Tau{\mathrm{T}}
\usepackage{graphicx}
\usepackage{xcolor,soul}
\usepackage{hyperref}
\usepackage{cite}
\usepackage{bm}
\usepackage{mathrsfs}  

\allowdisplaybreaks

\newtheorem{theorem}{Theorem}

\newtheorem{lemma}[theorem]{Lemma}
\newtheorem{proposition}[theorem]{Proposition}
\newtheorem{assumption}{Assumption}
\newtheorem{remark}{Remark} 
\newenvironment{proof}{\noindent{\bf Proof }}{\hfill$\Box$ \bigskip}

\newcommand{\minimize}[1]{\displaystyle\minim_{#1}}
\newcommand{\minim}{\mathop{\hbox{\rm minimize}}}
\newcommand{\maximize}[1]{\displaystyle\maxim_{#1}}
\newcommand{\maxim}{\mathop{\hbox{\rm maximize}}}
\newcommand{\sbjt}{\mathrm{subject\ to}}

\begin{document}
%\title{Optimal Retail Tariff Design with Prosumers: Balancing Economic Efficiency and Energy Equity }
\title{Optimal Retail Tariff Design with Prosumers: Pursuing Equity at the Expenses of Economic Efficiencies?}
%in an Integrated Wholesale and Retail Energy Market with Prosumers}

%\iffalse
\author{Yihsu~Chen,~%\IEEEmembership{Senior Member,~IEEE,}
        Andrew ~L. Liu, ~%\IEEEmembership{Member,~IEEE,}
        Makoto~Tanaka,~%\IEEEmembership{Member,~IEEE,}
       and~Ryuta~Takashima~%\IEEEmembership{Member,~IEEE \vspace{-1em}}% <-this % stops a space

%\thanks{$^*$Corresponding author: Y. Chen. Professor of Technology Management in Sustainability, Electrical, and Computer Engineering, Environmental Studies, University of California Santa Cruz, Santa Cruz, CA, United States. E-mail: \tt{yihsuchen@ucsc.edu.}}% <-this % stops a space
}

\maketitle

\begin{abstract}
Distributed renewable resources owned by prosumers can be an effective way of fortifying grid resilience and enhancing sustainability.
However, prosumers serve their own interests and their objectives are unlikely to align with that of society. 
This paper develops a bilevel model to study the optimal design of retail electricity tariffs considering the balance between economic efficiency and energy equity. 
The retail tariff entails a fixed charge and a volumetric charge tied to electricity usage to recover utilities' fixed costs.
We analyze solution properties of the bilevel problem and prove an optimal rate design, which is to use fixed charge to recover fixed costs and to balance energy equity among different income groups. This suggests that programs similar to CARE (California Alternative Rate of Energy), which offer lower retail rates to low-income households, are unlikely to be efficient, even if they are politically appealing. 
%We further present a numeric case study to illustrate how prosumers' incentives to sell/buy energy to/from the grid can be affected by different tariff designs under a net-billing policy. 
\end{abstract}

%\begin{IEEEkeywords}
%prosumers, energy expenditure incidence, 
%retail and wholesale market, 
%net-metering, net-billing, fixed cost recovery, bilevel optimization
%\end{IEEEkeywords}

% \fi
\IEEEpeerreviewmaketitle

\section{Introduction}

\IEEEPARstart{C}{oncerns} about climate change, resilience to hazardous events, and sustainability have shifted the electric power sector in the U.S. and elsewhere toward more involvement on the demand side to harness flexible distributed energy resources (DERs). More recently, the U.S. Federal Energy and Regulatory Commission (FERC) issued Order 2222, which removes barriers to the integration of DERs into wholesale electricity markets \cite{ferc2222}.
%breaking new ground toward creating the grid of the future . 
More specifically, the ruling allows the integration of multiple DERs owned by different entities with different sizes and diverse technologies to participate in the regional organized wholesale capacity, energy, and ancillary services markets alongside traditional resources.
This ruling  
%not only provides the opportunities for rethinking market design for the power sector but also challenges the conventional, top-down power grid architecture based on the supply side \cite{kristov19}.
%Furthermore, it 
offers incentives for households that own DERs (known as prosumers) to ``value-stack'' their assets to provide various types of energy-related commodities to the grid.
Already, we are witnessing some activities in the marketplace in response to or anticipation of the order. 
For instance, OhmConnect,  a clean tech company, recently announced a plan to link homes spread in California to form a 550 MW virtual power plant (VPP) of distributed energy resources. %\cite{OhmConnect20}.

Naturally, prosumers  are  self-served  for  their  own  interests,  and  their behavior  resulting  from  optimizing  their  private  benefits  is unlikely  to  be  in  the  best  interests  of  the  energy  market  as a  whole.
%\hl{Over reliance on the grid leads to over-production by conventional producers in the wholesale market, resulting in productive inefficiencies. (AL: what does this mean?)}
%\hl{Excessive selling of power to the grid by prosumers induces over self-generation or under self-consumption, thereby making themselves worse-off. (AL: is it incentive-compatible for prosumers to do that? If not, why would they do so?)}  
Not surprisingly, how to compensate for the energy produced by prosumers has emerged as a critical issue that can facilitate or impair the deployment of DERs and is currently subject to contentious debates\cite{Costello14,bushnell18,chakraborty19}.

The situation is also complicated by the existing formation of retail tariff. 
%lumpiness of fixed costs related to the network infrastructure for example, which need to be recovered by utilities in the current retail tariff, as well as other costs.
In general, a retail tariff consists of four core elements: (i) costs of electric energy; that is, wholesale locational marginal prices (LMPs), (ii) costs of other energy-related services, such as operating reserves or capacity costs, (iii) costs for network-related services, including investment and maintenance costs of transmission and distribution network assets, and (iv) charges to recover policy costs, such as procurement costs to support state's RPS (renewable portfolio standards) \cite{mit:16}.%\footnote{At least in California, the Public Utility Commission (PUC) has included a torrent of policy goals into electricity retail rate making, such as subsidies for rooftop solar and electrical vehicle charging stations, discount for low-income customers, energy efficiency programs, school water quality, and more recently, wildfire mitigation and compensation \cite{borenstein20}.}
%Some aspects of retail tariffs are worth noting.
%First, (ii) sometimes is referred to as going-forward costs; that is, the incremental costs in order to maintain the reliability of a system. 
%It is represented as a peak-coincident load charge. 
 The last two items, (iii) and (iv), are generally lumpy and non-convex as they are not directly tied to the level of energy consumption.
These two elements are dubbed as ``residual costs'' in  \cite{burger19}. 
The breakup of those four elements depends on specific markets; for example, the energy or LMP component can be as low as 10\% in the Netherlands or as high as 60\% in New Jersey \cite{mit:16}. 

\iffalse
Currently, there are two commonly implemented systems in the power sector: net-billing and net-metering. 
Whereas the net-metering system is more common in the US, the net-billing is more popular in Europe. 
Under a net-billing system, two meters are  installed,  recording  two  quantities,  i.e.,  power  withdrawn from and power injected into the grid, where withdrawal and injection  can  be  subject  to  difference  prices.  
In  contrast,  a single  meter  can  be  installed  under  a  net-metering  system  to record  both  transactions,  subjecting  them  to  the  same  price.
\fi

Recovering residual costs is a thorny political-economic endeavor, which may facilitate or impair the deployment of residential DERs, and has been subject to contentious debates \cite{Costello14,chakraborty19}.
%\cite{Costello14,bushnell18,chakraborty19}.
Two systems are of great interest. The first is referred to as net-metetering; that is, prosumers are only billed for their ``net" energy use. The energy they sell back to the grid will be paid at the same rate as buying from the grid. 
The second system is called net-billing, under which two meters are  installed,  recording  two  quantities:  energy withdrawn from and energy injected into the grid. The withdrawal and injection  can  be  subject  to  different prices. While the net-metering is the most common approach and has provided strong incentives for DER investment, it is also causing serious equity issues, as the more prosumers take advantage of net metering, the fewer residual costs are paid into the system, resulting in higher rates for non-net metering customers, likely those of low-income.

Recently, the California Public Utility Commission (CPUC) engaged in a regulatory process to revamp its net-metering policy,
%Not surprisingly, the process attracts considerably attentions from various stakeholders, especially from the solar industry who believes that the future of the industry is on the line.
as the CPUC is fully aware of its drawbacks.\footnote{\url{https://www.cpuc.ca.gov/-/media/cpuc-website/divisions/energy-division/documents/net-energy-metering-nem/nemrevisit/430903088.pdf}} 
%For example, in its rule-making, it states ``{\it ... we find the structure of the successor tariff should be revised to be a better version of, with an export compensation rate better aligned with the value exported energy provides to the grid...''} and {\it ``Hence, export compensation should be based on avoided cost values and successor tariff customers should pay for their usage of the grid.''}\footnote{\url{https://www.cpuc.ca.gov/-/media/cpuc-website/divisions/energy-division/documents/net-energy-metering-nem/nemrevisit/430903088.pdf}}
Such issues have also been vetted by the academics, which
%surrounding fixed cost recovery in the presence of DERs have also been vetted by the academics. 
have been described as ``revenue erosion'' or ``network defection''; that is, utilities are forced to increase the retail tariff to compensate for the revenue deficiency, further exacerbating the situation and leading to the so-called ``death spiral" \cite{Borenstein15,Picciariello15,Castaneda17,Kubli18,Darghouth16}.
Some empirical evidence emerges: for instance, using data from three investor-owned utilities in California, Wolak \cite{wolak18} finds that two-thirds of the increases in residential distribution prices can be attributed to the growth of solar capacity.

%A utility would prefer a monthly charge to recover their fixed costs because of revenue certainty.
%However, the fact that consumers face a lower volumetric retail price implies that they have less incentive to conserve energy.
%The urgent question the sector facing is how to price energy produced by prosumers so as to the market can benefit from their presence. 
A number of recent studies have also addressed the issues of pricing energy produced by prosumers.
%Abada et al. \cite{abada20} apply cooperative game theory to examine an individual's incentive to participate in energy communities.
%The study finds that the willingness to join depends on installation costs, the magnitude of aggregated benefits, coordination costs, and energy sharing rules, as the well as network tariff. 
Clastres et al. \cite{clastres19} estimate the extent of cross-subsidies between prosumers and conventional consumers in France. 
The authors also conclude that a demand charge may alleviate the network defection or death spiral problem facing distributed system operators. 
Using stylized models, Gautier \cite{Gautier18} concludes that net-metering decreases the payment from prosumers, which is cross-subsidized by the higher bills of conventional consumers.
%The system also leads to too many prosumers and provides no incentive to synchronize local production and consumption.
More recently, Gorman et al. \cite{gorman20} compare grid costs to off-grid costs of more than 2,000 utilities in the U.S. and find that network defection could increase from 1\% to 7\%, with 3\% in the Southwest region and California and 7\% in Hawaii.
\textcolor{black}{However, little attention was given to examine the impact of retail tariffs on the energy equality among different income groups in the presence of prosumers. 
An exception is our earlier work that compares the energy expenditure incidence among different income groups when prosumers are subject to a net-metering and a net-billing policy \cite{chen21}.
The paper concludes that  net-metering is more regressive than net-billing under the volumetric tariff. 
A hybrid policy, which also features an income-based fixed charge may potentially improve energy equity.}
The current study extends our previous work to offer a policy prescription on optimal retail tariff design in the face of a growing presence of prosumers.
The problem is formulated as a bilevel optimization problem: the upper level represents a public utility commission (PUC)'s decision-making problem that has to decide a certain retail tariff structure to guarantee the recover utility's fixed costs as well as maintaining energy equity. 
The lower level represents an market equilibrium that consists of prosumers, consumers, producers, and an independent system operator (ISO), with the prosumer/consumers' retail rates set by the upper-level PUC.
%need to be recovered from either a volumetric charge or a hybrid system that entails an additional fixed charge while power sales from prosumers are subject to a net-billing policy.
Although each prosumer may be relatively small, possessing limited ability to engage in the bulk energy market, we assume that an entity integrates a large number of prosumers and participates in the bulk energy market on their behalf; this is consistent with FERC Order 2222's requirements.
The prosumers are endowed with renewables and decide on the amounts of self-consumption, dispatchable energy to produce, such as from back-up generators or energy storage, and energy to sell to or buy from the bulk energy market to maximize their net benefit. 
%which is characterized by their ``retail'' inverse demand functions. 
The ISO minimizes the generation costs while treating sales or purchases by the prosumers as exogenous.

\textcolor{black}{Similar to the earlier work by Woo \cite{woo88}, we also explicitly consider the problem facing a PUC and distinguish the retail rate from the wholesale rate. 
However, unlike it, we extend the analysis to consider the energy expenditure incidence among income groups to address equity issues.}
Our analysis of the theoretical properties of the bilevel is also worth noting. More specifically, we prove that \emph{laissez-faire} is socially optimal; that is, zero volumetric charge will maximize the social surplus, as it will not distort the equilibrium price in the wholesale market, and energy equity can be achieved through different fixed charges among different income groups. While a fixed-charge-only tariff is implausible in reality, our formulation is amenable to computing a second-best solution when a proportion of the utility's fixed cost is required to be recovered from volumetric charges. 
\textcolor{black}{While our lower-level problem} is related to \cite{clastres19,Gautier18}, it is different in significant ways. 
In particular, we consider the transmission network and market details, \textcolor{black}{e.g., pool-type market settlement, capacity ownership, generation capacity constraints, retail-wholesale market linkage}, which are crucial in determining realistic electricity market outcomes. 
%Similar to \cite{chen21}, we also \hl{benchmark (AL: I'm not sure what `benchmark' means here. )} the case to 2015's RECS data by assuming an energy incidence of 1.5\% \textcolor{blue}{\hl{in order to discover income level by different groups (AL: I don't understand what this means. What does it mean to ``discover income level"?)}} and solve prosumers' load endogenously to examine its impact on energy incidence.
%\hl{\textcolor{blue}{Our interests lie in understanding how an optimal retail rate can lead to a market outcome that can ``restore'' to the benchmark case.} (AL: I don't understand this sentence either. What do you mean ``restore" to the benchmark case? What is the benchmark case, and what are to be restored: price, demand, supply?}

\iffalse
Third, we note that the size of the prosumers is exogenous since our interest is not the optimal proportion of prosumers.
\textcolor{red}{Because, in reality, any changes to the retail tariff is  a lengthy legal process, unlikely to be socially optimal, but is undoubtedly subject to various inputs from stakeholders with competing interests, e.g., utilities, the PUC, and consumer advocates.
Thus, our intention herein is to understand how an income-based fixed-cost charge may improve the equity in energy expenditure among different income groups. 
}
\fi
%However, from our results, we are able to infer that the net-metering policy is always likely to lead to too many prosumers since it generously over-compensates prosumers for their renewables and backup generation.
%This finding is consistent with \cite{Gautier18}. 

The remainder of this paper is organized as follows.
%Section \ref{sec:rev} reviews the relevant literature.
Section \ref{sec:mod} presents the lower- and upper-level models of the bilevel problem. Solution properties and theoretical results are shown in Section \ref{sec:Theo}. 
%Section \ref{sec:analysis} analytically compares the net-metering and first-best cases. 
A numerical case study is presented in Section \ref{sec:case}.
Finally, concluding remarks are provided in Section \ref{sec:con}.

%In principle, fixed cost can recover from volumetric 

%\cite{borenstein20}

%{\color{black} need more ...}

%\newpage

\section{Model}\label{sec:mod}
We present the complete model in this section, starting with the lower-level market equilibrium formulation, followed by the upper-level problem to maximize social surplus and energy equity. 
The resulting problem can written as either a mathematical program with equilibrium constraints (MPECs) or a bilevel problem (BLP), with the former formulation amenable to computation and the latter one easier for theoretical analysis. \vspace{-1em} 

\subsection{Lower-Level Problem}
The lower-level problem consists of problems faced by the consumers, prosumers, power plants, and the ISO. Throughout the paper, we make the blanket assumption that the market is perfectly competitive; that is, all market participants are price-takers of the market prices, without contemplating on how to manipulate the equilibrium prices through their unilateral actions. While market power abuse has always been a concern for wholesale energy markets, retail ratemaking usually lasts for a certain period of time; that is, retail rates do not change frequently. It is therefore unreasonable to assume that a wholesale market is subject to sustained market power abuse.

\subsubsection{Consumers}
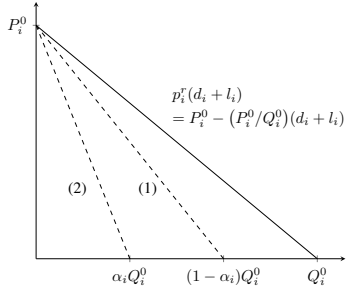
\begin{figure}[!htbp]
\centering
\begin{tikzpicture}[scale=0.6]
\begin{axis}[axis lines=center, xmin=0, xmax=22, ymin=0, ymax=22, xtick ={20/3,40/3,20}, xticklabels ={$\alpha_i Q^0_i$,$(1-\alpha_i) Q^0_i$,$Q^0_i$}, ytick ={20}, yticklabels ={$P^0_i$}]
\addplot[domain=0:20]{ -x+20};
\addplot[dashed,domain=0:20]{ -3*x+20};
\addplot[dashed,domain=0:20]{ -1.5*x+20};
\node[left] at (axis cs:4,6) {(2)};
\node[left] at (axis cs:9,6) {(1)};
\node[right] at (axis cs:9.3,14) {$p_i^{r}(d_i+l_i)$};
\node[right] at (axis cs:9.3,12) {$=P^0_i - \big({P^0_i}/{Q^0_i}\big)(d_i+l_i)$};
\end{axis}
\end{tikzpicture}%\vspace{-1em}
\caption{An illustration of retail demand curves}
\label{fig:demand}      
\end{figure}

Consider an energy market that has $N$ nodes and $K$ transmission lines that connect the nodes. Consumers at each node $i = 1, \ldots, N$ 
are grouped into two types, including conventional consumers and prosumers, whose marginal benefit functions (that is, their willingness-to-pay functions), denoted by $p_i^{con}$ and $p^{pro}_i$, respectively, are represented by the following linear inverse demand functions:
\begin{align}
p^{con}_i(d_i)& =P^0_i - \Big({P^0_i}/{\big((1-\alpha_i)Q^0_i\big)}\Big)d_i,\  \forall\ i \label{demand}\\
%\end{align}
%\begin{align}
p_i^{pro}(l_i)& =P^0_i - \Big({P^0_i}/{\big(\alpha_i Q^0_i\big)}\Big)l_i,\ \forall\ i, \label{demandp}
\end{align}
where $P^0_i > 0$ and $Q^0_i >0$ respectively represent the vertical and horizontal intercepts of the ``horizontally aggregated'' retail inverse demand function: \textcolor{black}{$p_i^{r}(d_i+l_i)=P^0_i - \big({P^0_i}/{Q^0_i}\big)(d_i+l_i)$, as illustrated in Fig. \ref{fig:demand}}.
The quantities demanded by conventional consumers and prosumers are denoted by $d_i$ and $l_i$, respectively.
The parameter $\alpha_i$ is the fraction of prosumers at node $i$. 
Note that while $\alpha_i$ varies between 0 and 1, the aggregated demand does not change. 
%\vspace{-5pt}

Let $p_i$ denote the wholesale energy price at node $i$, and $\tau^b$ be the volumetric charge of energy purchase for all consumers/prosumers, which is a part of consumers' retail rates.\footnote{Note that retail rates are usually the same covering a broad area of customers, and hence, we do not have a node index of $\tau$, but can certainly do so.} The other part of the retail rates is the fixed charge. Since the fixed-charge rate will serve as a main tool to realize energy equity, we assume that conventional consumers and prosumers can be subject to different fixed-charge rates, and denote them as  $\phi_i^{con}$ and $\phi_i^{pro}$, respectively.

%based on the fixed network or other types of stranded costs, such as renewable contracts, that need to be recovered. 
%Hence, consumers' per-unit electricity rate is 
%Thus, $p_i=p^{con}_i - \tau$ holds, where $p_i$ is the wholesale power price in node $i$. \textcolor{black}{In addition to a volumetric charge $\tau$, customers in node $i$ are subject to an income-dependent fixed charge $\phi_i$, of which $(1-\alpha_i) \phi_i$ and $\alpha_i \phi_i$ are borne by conventional consumers and prosumers, respectively.}
With the marginal benefits and costs defined, conventional consumers at each node $i = 1, \ldots, N$ maximize their net benefits (also referred to as surplus) by solving:
\begin{maxi!}[3]
	{d_i\geq 0}{\int^{d_{i}}_{0} p^{con}_i(m_i)dm_i- (p_i + \tau^b)d_i - \phi_i^{con}.}{\label{eq:con}}{} \tag{3}
\end{maxi!}
Since $p_i$, $\tau^b$, and $\phi^{con}_i$ are all exogenous to consumers, the optimization problem is easily seen to be a strongly convex problem with the given linear inverse demand function as in\eqref{demand}. 
Hence, an optimal solution always exists with respect to any ($p_i$, $\tau^b$, $\phi^{con}_i$), and the first-order optimality conditions, aka the KKT conditions, are both necessary and sufficient for optimality. The collection of conventional consumers' KKT conditions are that for $i = 1, \ldots, N$: 
%\begin{subequations}
\begin{align}
%\label{8a}
0 \leq d_{i} \perp  P_i^0 - \Big({P^0_i}/{\big((1-\alpha_i)Q^0_i\big)}\Big)d_i - (p_i + \tau^b) \leq 0,  \label{eq:d_KKT}
\end{align}
where the `$\perp$' sign means that the product of the scalars or vectors is 0, 
and such a constraint is referred to as a complementarity constraint.
The KKT conditions have intuitive economic interpretation: at an optimal solution (denoted with a `$*$' superscript), if $d_i^* > 0$, then consumers choose to purchase energy at the level where the marginal benefit, $P^0_i - \Big({P^0_i}/{\big((1-\alpha_i)Q^0_i\big)}\Big)d^*_i$ equals the marginal cost, which is the retail price $p^r_i:= p_i+\tau^b$. 
%\end{subequations}

%We describe their optimization in this section. 

%¥noindent {¥bf{Prosumers}}
%\subsection{Prosumers}
\subsubsection{Prosumers}\label{subsec:Prosumer} 
For prosumers, they pay the same volumetric charge $\tau^b$ when buying from the grid. However, when they sell to the grid, we assume that the rate they receive is $p_i + \tau^s$. If $\tau^s = \tau^b$, then it is the net-metering policy; otherwise, it is net billing. Note that while $\tau^b$ is always non-negative, $\tau^s$ can be positive or negative. When $\tau^s>0$, the prosumers effectively receive a ``subsidy'' in addition to the wholesale price $p_i$.
In the case where $\tau^s<0$, it means that the prosumers are subject to a ``tax'' %(e.g., transmission tariff) 
when selling power to the grid.%\footnote{It could also be other types of fixed or sunk costs that a utility attempts to recover, e.g., renewable procurement cost, those associated with discounts to low-income households, legacy stranded costs, or even drinking water related projects.
%However, not all sunk costs are recoverable: the capital employed or equipment needs to be “used and useful” and “prudently operated.” See Hempling \cite{hempling15} for more discussion.}
%with a negative $\tau^s$ representing the policy such that prosumers at a node $i$ will only get paid of the wholesale price $p_i$ minus a volumetric charge, such as the cost of accessing the distribution network. 

With the exogenous volumetric rates $\tau^b$, $\tau^s$ and fixed rate $\phi^{pro}$, we posit that a prosumer maximizes its surplus by deciding i) energy to buy ($z_i^b$) from or sell ($z_i^s$) to \textcolor{black}{the grid at} %producer $f$ in
node $i$,
%\footnote{As the equivalence between a power market based on pool-type transactions and on bi-lateral contracts have been alluded to in \cite{HobbsLinearcomplementaritymodels2001}, we believe that our assumption herein is reasonable and can be seen as an extension.} 
ii) consumption level $l_i$ given renewable output $R_i>0$, and iii) generation $g_i$ from the backup dispatchable technology with a cost $C^g_i(g_i)$.
%\footnote{ \hl{One example (AL: an example of what, with backup generation? If so, can you specify the types/fuel of the back-up generation?)} is a company that acts as an aggregator managing more than 280 commercial, industrial, and residential customers operating in CAISO (California Independent System Operator) to provide energy and frequency regulation among other services \cite{burger19}.}
The prosumer's problem at node $i$ is: \vspace{-1em}
\begin{subequations}
\begin{align}
	\maximize{z^b_{i},z^s_{i},l_i,g_i \geq 0 }\ & \ (p_i+\tau^s)z^s_{i}-\left(p_i+\tau^b
	\right)z^b_{i}\nonumber \\[-5pt]
	& +\int^{l_{i}}_{0} p^{pro}_i(m_i)dm_i- C^g_i(g_i)- \phi^{pro}_i \label{eq:nb-obj} \\
\mathrm{subject\ to} & \ l_i + z^s_{i}-z^b_{i} - g_i - R_i = 0 \quad (\delta_i) \label{eq:nb-con1} \\ 
  &\ g_i \leq G_i \quad (\kappa_i)\label{eq:nb-con2}
%\addConstraint{0 \leq z_i^s}{\perp z_i^b \geq 0}{(\nu^s_i,\ \nu_i^b)}{\label{eq:ProsumerComp}}
% & \  l_i,\ g_i,\ z_i^s,\ z_i^b \geq 0. \label{eq:Prosumer_Nonneg} 
	%\addConstraint{z^s_i}{\leq Z_i}{(\gamma_i)}{\label{eq:nb-con3}}
\end{align}
\end{subequations}
%As specified in the objective function \eqref{eq:nb-obj},  the power withdrawn from the grid is priced at \textcolor{black}{ $p_i+\tau^b$}, where $\tau^b$ is always positive 
%\textcolor{black}{as a usual transmission tariff for end-users}.
% \textcolor{black}{The power injected into the grid is compensated by $p_i+\tau^s$, where $\tau^s$ can be either positive or negative. }%\footnote{\textcolor{green}{When $\tau^s=0$, a net-billing system is equivalent to a net-metering scheme.}}
 In the constraint set, Eq. (\ref{eq:nb-con1}) ensures that \textcolor{black}{the net} demand, $l_i + z^s_{i}-z^b_{i}$, is balanced with the prosumer's own renewable and backup generation ($R_i + g_i$). 
Eq. (\ref{eq:nb-con2}) limits the backup output to be less than its capacity $G_i$. 

Two things to note about the above optimization problem. First, if $\tau^s >  \tau^b$, the optimization problem is clearly unbounded.
%as both $z_i^s$ and $z_i^b$ can reach $+\infty$ while satisfying all constraints, and the value of the objective function will also approach $+\infty$. 
This is intuitive: if $p_i + \tau^s > p_i + \tau^b$, meaning that selling electricity back to the grid earns more than buying from the grid. Then a prosumer can simply arbitrage and earn an infinite amount of profit. To rule out this case, we make the no-arbitrage assumption that $\tau^s \leq \tau^b$.
%by buying an infinite amount of energy from the utility and immediately selling back,

Second, under the no-arbitrage assumption, for the buy and sell decisions, $z_i^s$ and $z_i^b$ should not both be positive in an optimal solution, from a common sense perspective. 
Mathematically, however, this is not guaranteed, unless we have some explicit constraints such as $z_i^s \cdot z_i^b = 0$. In the following, we present the simple fact that such an explicit constraint is not necessary. To do so, we first write down the KKT conditions of the prosumers optimization problems (assuming that $C_i^g(\cdot)$ is differentiable). 
\begin{subequations}
	\begin{align}
		%\label{8a}
		& 0 \leq l_i \perp P^0_i-\frac{P^0_i}{\alpha_i Q_i^0}l_i - \delta_i \leq 0,\ \forall i \label{eq:ProKKT_l}\\
		& 0 \leq g_i  \perp - C^{g\prime}_i(g_i)  + \delta_i -\kappa_i  \leq 0,\ \forall i \\
		& 0 \leq \kappa_i \perp g_i - G_i \leq 0,\ \forall i\\  
		& 0 \leq z_i^s \perp (p_i + \tau^s) - \delta_i  \leq 0,\ \forall i    \label{eq:Pro_zs}\\ %\tag{3a*}
		& 0 \leq z_i^b \perp - (p_i + \tau^b)  + \delta_i \leq 0,\ \forall i	 \label{eq:Pro_zb}\\
		& \delta_i \ \mathrm{free},\  l_i + z^s_{i}-z^b_{i} - g_i - R_i = 0,\ \forall i. \label{eq:ProKKT_delta}
	\end{align}
\end{subequations}
Since the constraints \eqref{eq:nb-con1} -- \eqref{eq:nb-con2} are all linear, and with a given tuple $(p_i, \tau^s, \tau^b, \phi_i^{pro})$,  the objective function is concave (under the assumption that $C_i^g(\cdot)$ is a convex function), the KKT conditions are again necessary and sufficient optimality conditions. Then we have the following result.
\begin{lemma}\label{lem:zszb}
Assume that $C_i^g(\cdot)$ is continuously differentiable and convex. 
With a given tuple $(p_i, \tau^s, \tau^b, \phi_i^{pro}) \in  \Re_+ \times \Re \times \Re_+ \times \Re$ in which $ \tau^s \leq  \tau^b$,  an optimal solution of the problem \eqref{eq:nb-obj} -- \eqref{eq:nb-con2}, denoted by $(l^*_i,g^*_i,z^{s^*}_{i},z^{b^*}_{i})$, exists. In addition, if $\tau^s \neq \tau^b$, then $z^{s^*}_{i} \cdot z^{b^*}_{i} = 0$.
\end{lemma}
\begin{proof}
	Let $X_i^{pro} \in \Re^4$ denote the feasible region of the prosumers' problem at node $i$. It is easy to see that $X_i^{pro} \neq \emptyset$ since the zero vector is always in $X_i^{pro}$. $X_i^{pro}$ is also clearly a closed set. When $\tau^s \leq \tau^b$, the objective function \eqref{eq:nb-obj} goes to $-\infty$ for any $(l_i,g_i,z^{s}_{i},z^{b}_{i}) \in X_i^{pro}$ with $|| (l_i,g_i,z^{s}_{i},z^{b}_{i})|| \rightarrow \infty$, which means that \eqref{eq:nb-obj} is coercive on $X_i^{pro}$. Since \eqref{eq:nb-obj} is also continuous, an optimal solution exists by a variant of the well-known Weierstrass' Theorem (such as Proposition A.8 in \cite{bertsekas99}). 
	
	For the second part, when $\tau^s \neq \tau^b$, if $z^{s^*}_{i} \cdot z^{b^*}_{i} > 0$, the conditions \eqref{eq:Pro_zs} and \eqref{eq:Pro_zb} imply that $p_i + \tau^s = \delta_i^* = p_i + \tau^b$, a contradiction.
\end{proof} 

When $\tau^s = \tau^b$, clearly the optimal solutions $z^{s^*}_{i}$ and $z^{b^*}_{i}$ are not unique and can be both positive. In this case, we can simply define $\tilde{z}_i^s := \max\{z^{s^*}_{i} - z^{b^*}_{i}, 0\}$ and $\tilde{z}_i^b := \max\{z^{b^*}_{i} - z^{s^*}_{i}, 0\}$; then at most one of them is nonzero in an optimal solution. In the following sections, we will see that it is always the net of a prosumer's decision, namely, 
$z^{s}_{i} - z^{b}_{i}$, that appears in the other part of the market equilibrium model. Hence, not including an explicit constraint such as $z_i^s \cdot z_i^b = 0$ will not affect the outcomes of a market equilibrium in any way, and omitting such combinatorial constraints will considerably simplify both theoretical analysis and computation.

%This is implied by the violation of the MFCQ, since LICQ there is an 

%In Section \ref{sec:Equilibrium} we will discuss the relationship of an optimal solution of the prosumer's problem and the set of stationarity conditions 

%¥noindent {¥bf{The Independent System Operator}}

\subsubsection{Power Producers and The ISO} We assume that
%Under the assumption of perfect competition, it is well known that a market equilibrium maximizes the social surplus. Hence, instead of modeling individual power producers' bids into an ISO-operated wholesale market, we combine producers and an ISO into one cost minimization problem. Coupled with the previous consumers/prosumers' models, they together yield the market equilibrium
an ISO collects offers from electric power producers to minimize the total generation cost, while treating the demand and prosumers' buy and sell decisions as exogenous. The optimization problem is as follows:
\vspace*{-5pt}
\begin{subequations}
\begin{align}
	\minimize{g_{ih}\geq 0,\ y_{i}\ \mathrm{free}}\ & \ \sum_{i=1}^N\sum_{h \in H_{i}}  C_{ih}(g_{ih}) & \label{eq:iso} \\
\mathrm{subject\ to} & \ g_{ih} - G_{ih}\leq 0 \hspace*{10pt} (\rho_{ih}),\ \forall i, h \in H_i \label{eq:iso_con1}	\\
%\addConstraint{y_{i} - \sum_{h \in H_{i}} g_{ih}-z_i+b_i + d_{i}}{=0}{\qquad(p_i)}{\label{eq:iso_con2}}
	& \ \sum_{i=1}^N y_{i}= 0 \hspace*{25pt} (\theta) \label{eq:sumY=0} \\
	& \ \sum_{i=1}^N PTDF_{ki} y_i \leq T_k \hspace*{18pt} (\lambda_k^+),\ \forall k \label{eq:iso_con4}\\
	& \ -\sum_{i=1}^N PTDF_{ki} y_i \leq T_k \hspace*{10pt} (\lambda_k^-),\ \forall k \label{eq:iso_con5} \\
    & \ y_i-\sum_{h \in H_i}g_{ih}-(z^s_i-z^b_i) +d_i =0 \hspace*{10pt} (p_i),\ \forall i \label{eq:nb-price} \vspace{-2em}
% & \ g_{ih} \geq 0,\ y_i\ \mathrm{free} \quad \quad \forall i, h \in H_i. \label{eq:ISO_nonneg} 
\end{align}
\end{subequations}
In the above problem, \eqref{eq:iso_con1} is the generation capacity constraint. The set $H_i$ represents all power plants at node $i$; therefore, we do not need to assume that there is only one power plant at each node $i$.
Eq. (\ref{eq:sumY=0}) ensures that the total net injection/withdrawal in the system is equal to zero, where $y_i$ represents the energy flow from an arbitrarily assigned hub node to node $i$. We represent the transmission network as a hub-spoke system; that is, the energy flows from node $i$ to $j$ are considered as from $i$ to the hub, and from the hub to $j$. 
Eqs. (\ref{eq:iso_con4})--(\ref{eq:iso_con5}) describe that the flow in link $k$ is less than its transmission limit $T_k$.
The term $PTDF_{ki}$ represents the power transfer distribution factors based on linearized DC flows. Eq. \eqref{eq:nb-price} is the mass balance constraint, whose shadow price, denoted by $p_i$, is exactly the wholesale energy price at node $i$ (that is, the LMP at $i$). 

%\addConstraint{\sum_f (s_{i} + b_{i} - z_{i}) - \sum_{f, h \in H_{fi}}x_{fih}}{=}{y_i}{}{\qquad (p_i)}
	%\addConstraint{(\ref{eq:nb-con1}) \quad \text{or} \quad %(\ref{eq:nm-con1})}{}{\qquad (p_i)},
	%\addConstraint{z_{fi},b_{fi},l_i,g_i}{\geq 0}.{\label{1e}}\nonumber
%This constraint is case-dependent, that is, (\ref{eq:nm-price}) and (\ref{eq:nb-price}), representing the net-metering and the net-billing case, respectively. 

%
%\begin{align}
%& y_i - \sum_{h \in H_i}g_{ih} -z_i + d_i=0  \qquad (p_i),\ \forall i, 
%\label{eq:nm-price}  \\
%& y_i-\sum_{h \in H_i}g_{ih}-(z^s_i-z^b_i) +d_i=0 \qquad (p_i),\ \forall i \label{eq:nb-price} \tag{7f}
%\end{align}
%\textcolor{red}{The condition shifts the demand by conventional consumers in the wholesale market operated by the ISO, given the decision of the prosumers. For example, if the prosumers' net position is to sell or  $z^s_i-z^b_i>0$ in node $i$, the effective “wholesale” demand is then reduced or shifts to the left by $z^s_i-z^b_i$ through (\ref{eq:nb-price}). 
%Similarly, if the prosumers' net position is to purchase or $z^s_i-z^b_i <0$ from the grid, the wholesale demand increases or shifts to the right, reflecting the demand from both the conventional consumers and the prosumers.}

To aid in model development and analysis, we make the following blanket assumption throughout the paper. 
\begin{assumption}\label{as:QuadraticCost}
The generation cost function $C_{ih}(\cdot)$ is
%\begin{equation}\label{eq:QuadraticCost}
$C_{ih}(g_{ih}) = a_{ih}g_{ih} + \frac{1}{2}A_{ih}g_{ih}^2$, 
%\end{equation}
with the input parameters $a_{ih} > 0$  and $A_{ih} > 0$ for all $i = 1, \ldots, N$ and $h\in H_i$.
\end{assumption}

According to Eq. \eqref{eq:nb-price}, with a given $d_i$, 
the variable $y_i$ is implicitly bounded for $i = 1, \ldots, N$, since $g_{ih}$'s are bounded, and so is the quantity $z_i^s - z_i^b$ based on Eq. \eqref{eq:nb-con1}. Hence, with Assumption \ref{as:QuadraticCost}, 
if the feasible region is not empty with respect to a given demand vector $(d_i)_{i=1}^N$ and $(z_i^b, z_i^s)_{i=1}^N$, then an optimal solution of the above optimization problem exists, and the KKT conditions are necessary and sufficient of optimality, due to the all-linear constraints and the convex objective function. The detailed KKT conditions are: 
\begin{subequations}\label{e10}
\begin{align}
%& 0 \leq d_i \perp P^0_i-\frac{P^0_i}{(1-\alpha_i) Q^0_i}d_i-(p_i +\tau) \leq 0, \forall i \label{eq:foc-nm-d} \\ 
%& 0 \leq d_i \perp P^0_i-\frac{P^0_i}{(1-\alpha_i) Q^0_i}d_i-(p_i +\tau^b) \leq 0, \forall i \label{eq:foc-nb-d} \tag{6a*} \\
& 0 \leq g_{ih} \perp -C_{ih}^\prime(g_{ih})-\rho_{ih}+p_i \leq 0,\ \forall i, h \in H_i \label{eq:foc-iso-x}\\
& -\theta + \sum_{k=1}^K PTDF_{ki}(\lambda_{k}^- - \lambda_{k}^+) - p_i= 0,\ \forall i \\
& \textcolor{black}{0 \leq \rho_{ih} \perp g_{ih} - G_{ih} \leq 0,\ \forall i, h \in H_i} \\
& \theta\ \mathrm{free},\  \textcolor{black}{\sum_{i=1}^N y_{i}=0} \\
&0 \leq \lambda_{k}^+ \perp \sum_{i=1}^N PTDF_{ki}y_{i} - T_{k} \leq 0, \ \forall k \\
&0 \leq \lambda_{k}^- \perp -\sum_{i=1}^N PTDF_{ki}y_{i} - T_{k} \leq 0, \ \forall k \label{eq:foc-iso-f}\\
%& (\ref{eq:nb-price}) {\nonumber}.
& p_i\ \mathrm{free},\ 
y_i-\sum_{h \in H_i}g_{ih}-(z^s_i-z^b_i) +d_i=0,\  \forall i. \label{eq:foc-iso-g}
%& (\ref{eq:nm-price}) \qquad \text{or} \qquad (\ref{eq:nb-price}).\notag
%\sum_f (s_{fi} + b_{fi} - z_{fi}) - \sum_{f, h \in H_{fi}}x_{fih} = y_i, \forall i
\end{align}
\end{subequations}
%Eqs. (\ref{eq:foc-nm-d}) and (\ref{eq:foc-nb-d}) state that conventional consumers evaluate its marginal benefit against the retail price, that is, the wholesale price $p_i$ and the transmission % or grid 
%charge $\tau$ or $\tau^b$, when deciding their quantity demand, $d_i$.

%\vspace{0.5cm}
%¥noindent {¥bf{Fixed Cost Recovery}} 

\vspace{-1em}
\subsection{Upper-Level Problem}

\subsubsection{Fixed Cost Recovery}
First and foremost, revenue accrued through retail electricity rates (both volumetric and fixed rates) must cover utilities' fixed costs, which is ensured by the following constraint in the upper-level problem:
\vspace{-5pt}
\begin{align}
%\sum_i \big(\textcolor{black}{-} z_i +d_i\big)\tau+\sum_i \phi_i = \Tau
%\label{eq:nm}\\
\sum_{i=1}^N \big(\textcolor{black}{-z_i^s\tau^s}  +  z_i^b\tau^b +d_i\tau^b +\sum_{j \in {pro,con}} \phi^j_i \big)= \Tau,
\label{eq:nb}%\tag{8*}
\end{align}
where $\Tau$ represents the fixed cost to be recovered, which is exogenous to the model, and can be decided by utilities and approved by an energy commissioner.  
%While each participant's optimization problem represents its behavior in the power market,
%(\ref{eq:nb}) helps determine the transmission \textcolor{black}{charge} %, $\tau$, to reimburse to the transmission owners for their revenue adequacy $\Tau$ for net-billing and net-metering, respectively.
%\textcolor{red}{Thus, the term $\sum_{i, j \in {pro,con}} \phi^j_i$ %represents non-volumentric fixed charge to recover fixed costs.
%Note that $\Tau$ is a parameter, a similar setup as \cite{avau21}.}

\subsubsection{Equity and Energy Expenditure Incidence}
The upper-level decision-maker aims to strike the balance between maximizing the social surplus and maintaining energy equity when deciding on the retail electricity rates. To establish a model to aid decision making, we need quantitative measures of energy equity. In this work, we use the same measures (but with a small update) as developed in our earlier work \cite{chen21}. To make this paper self-contained, we reintroduce the measures here. 
The key concept is the energy expenditure incidence (EEI), defined for conventional consumers ($inc^{con}_i$) and prosumers ($inc^{pro}_i$) at node $i$ as follows:
\begin{align} 
	inc^{con}_i & =  \frac{(p_i + \tau^b)d_i+\phi^{con}_i}{\textcolor{black}{I^{con}_i}}, \label{eq:inc_con} \\
	inc^{pro}_i & =  \frac{(p_i + \tau^b)z^b_i+ \phi^{pro}_i+C_i(g_i) + SC_i}{\textcolor{black}{I^{pro}_i}}.  \label{eq:inc_pro}
\end{align} 
The EEI measures the proportion of consumers' spending on electricity compared to total household income (denoted by $I^{con}_i$ and $I^{pro}_i$ for conventional consumers and prosumers, respectively).  
The conventional consumers' energy spending is easy to understand, which is the sum of the fixed charge $\phi_i^{con}$ and the cost of purchasing $d_i$ energy at the rate of $p_i + \tau^b$; the prosumers' spending includes the cost of backup generation ($C_i(g_i)$) and the sunk costs $(SC_i)$ spent on purchasing or renting renewable energy equipment. Note that we do not subtract the prosumers' earnings (i.e., $(p_i + \tau^s)z_i^s$) from the energy spending in the numerator (or add it to $I^{pro}_i$ in the denominator) in \eqref{eq:inc_pro}, as we believe that prosumers should not be penalized for selling energy to the grid.  

With the EEI defined, our idea of energy equity is to minimize the differences in EEI between conventional consumers and prosummers, that is, to minimize $\sum_{i}^N(inc^{con}_i-inc^{pro}_i)^2$. The definition of EEI in \eqref{eq:inc_con} and \eqref{eq:inc_pro} involves both lower-level variables $(d_i, z^b_i, g_i, p_i)_{i=1}^N$ and upper-level variables $(\tau^b, \bm{\phi}^{con}, \bm{\phi}^{pro})$, where $\bm{\phi}^{con} := (\phi_1^{con}, \ldots, \phi_N^{con})\in\Re^N$ and $\bm{\phi}^{pro} := (\phi_1^{pro}, \ldots, \phi_N^{pro})\in\Re^N$. To simplify the argument, we define a vector $\bm{\zeta}\equiv(\tau^b, \tau^s, \bm{\phi}^{con}, \bm{\phi}^{pro}) \in \Re^{2N + 2}$, 
%and let $\bm{\zeta}^+$ be the vector without $\tau^s$. 
and use the notation $d(\bm{\zeta})$ to denote the optimal solution mapping with a given $\bm{\zeta}$; that is, $d(\bm{\zeta})$ is a set of all optimal solutions of the consumers' optimization problem \eqref{eq:con} with a given $\bm{\zeta}$. The notation for all other optimal solution mappings is the same. Then we define the difference-of-incidence function as:
\begin{equation}\label{eq:B_func}
B\left[\bm{\zeta};d(\bm{\zeta}), z^b(\bm{\zeta}), g(\bm{\zeta}), p(\bm{\zeta})\right] := \sum_{i}^N\left(inc^{con}_i-inc^{pro}_i\right)^2.
\end{equation}

%The parameter $M$ is a large positive number that represents a penalty.  
%The energy expenditure incidence in the current context, denoted by $inc$, is defined as the proportion of a consumer's income that is used to pay for electricity.
%(\ref{eq:inc_con}) and (\ref{eq:inc_pro}) define the incidence for conventional consumers and prosumers at location $i$, respectively:  

%\noindent where $p^r_id_i$ and $\phi^{con}_i$ denote the retail load payment and the fixed charge borne by conventional consumers, respectively. 
%The denominator $\textcolor{black}{I^{con}_i}$ gives the total incomes of these consumers. 
%The incidence of prosumers in node $i$ is defined as follows:

%noindent \textcolor{red}{where it only accounts for the costs of purchasing power from the network.
%That is, revenues from selling power into the grid cannot offset the %expenses and lower the incidence.
%This effectively bounds the incidence from the blow so that it is non-negative.
%}
%The terms, $\phi^{pro}_i$, $C_i(g_i)$, and $\textcolor{black}{I^{pro}_i}$ represent prosumers' fixed charge, costs of self-generation, and \textcolor{black}{total incomes, respectively.}

\subsubsection{The Complete Model}
As the upper-level decision maker does not want to distort wholesale energy markets with their ratemaking, we include the social-surplus maximization problem for the wholesale energy market in the upper-level problem as well. Social surplus of the wholesale market is defined as the sum of all the market participants' surplus, which we denote as $\Pi(\bm{x}; (\tau^b, \tau^s))$, with $\bm{x}$ representing the collection of all lower-level variables: $(d, l, z^s, z^b, g, y)$: 
\begin{align}
  \Pi[\bm{x}; (\tau^b, \tau^s)]& :=      \sum_{i}^N\left[\int_{0}^{d_i}p^{con}_i(m_i)\, dm_i - \tau^b d_i  \right] + \nonumber \\
  & \sum_i^N\Bigg[\tau^s z^s_{i} - \tau^b z_i^b  + \int_{0}^{l_i}p^{pro}_i(m_i)\, dm_i- C^g_i(g_i) \Bigg] \nonumber \\
  &	-  \sum_{i=1}^N\sum_{h \in H_{i}} C_{ih}(g_{ih}).  \label{eq:PI}
\end{align}
Note that by following the convention in economics, we do not include fixed charges $\bm{\phi}^{con}$ and $\bm{\phi}^{pro}$ in the social surplus, as the behavior of wholesale market participants is not affected by fixed charges in anyway. 
%\textcolor{blue}{Note that the terms $\sum_{i=1}^N \big(- \tau^b d_i +\tau^s z^s_{i} - \tau^b z_i^b -\sum_{j \in {pro,con}} \phi^j_i \big)$ are equal to the fixed constant $-\Tau$ in equilibrium because of Eq. \eqref{eq:nb}. XXX Makoto: we need to include "all" those terms or exclude "all" of them. But we cannot exclude only $\bm{\phi}^{con}$ and $\bm{\phi}^{pro}$ mathematically.XXX}
%\textcolor{red}{Note that fixed charges $\bm{\phi}^{con}$ and $\bm{\phi}^{pro}$ should have appeared in (\ref{eq:PI}) when calculating the social surplus . However, as they do not affect consumers' consumption levels and thus on the wholesale energy market, we exclude them.} 
%\hl{Note that fixed charges $\bm{\phi}^{con}$ and $\bm{\phi}^{pro}$ do not appear in the social surplus, as they do not depend on consumers' consumption levels and therefore do not have any impacts on the wholesale energy market.} 

With the social surplus and EEI-difference function defined, the complete two-level problem can be written as follows:
%The set of decision variables includes $\tau^b$, $\tau^s$, and $\phi_i$, as well as $\Omega :=\{d_i,l_i,g_{ih},g_i,z^s_i, z_i^b,y_i,\delta_i,\kappa_i, p_i, \lambda^+_i,\lambda^-_i, \theta\}$, where each variable within $\Omega$ can implicitly be represented as a function of $\tau^s$, $\tau^b$, and $\phi_i$. 
\begin{align}
    \maximize{\bm{\zeta},\  \bm{x}}\  & \ \Pi\left[\bm{x}; (\tau^b, \tau^s)\right] - M\cdot B(\bm{\zeta}; \bm{x}) \label{eq:MPEC}\\
    \sbjt & \  (\ref{eq:d_KKT}), (\ref{eq:ProKKT_l})-(\ref{eq:ProKKT_delta}),  (\ref{eq:foc-iso-x})-(\ref{eq:foc-iso-g})\text{, and } (\ref{eq:nb}), \nonumber \\
    & \tau^s \leq \tau^b, \ \tau^b,  \bm{\phi}^{con}, \bm{\phi}^{pro} \geq 0, \tau^s \  \mathrm{free}, \nonumber
\end{align}
where the parameter $M > 0$ in the objective function is to balance between the two (possibly conflicting) objectives of maximizing the social surplus and minimizing the difference of energy expenditure incidence. Constraint (\ref{eq:nb}) is the upper-level constraint to ensure revenue adequacy for utilities; while the constraints  (\ref{eq:d_KKT}), (\ref{eq:ProKKT_l})-(\ref{eq:ProKKT_delta}),  (\ref{eq:foc-iso-x})-(\ref{eq:foc-iso-g}), the collection of the KKT systems, represent the market equilibrium of the wholesale energy market with a given retail rate $(\tau^b, \tau^s)$.  

Problem \eqref{eq:MPEC} is an MPEC, which can be solved by nonlinear programming (NLP) solvers capable of dealing with complementarity constraints, such as KNITRO \cite{Knitro} or FILTER \cite{Filter}. Granted that \eqref{eq:MPEC} is a nonconvex problem, only a locally optimal (or stationary) solution can be computed using an NLP solver. 
%Directly computing a globally optimal solution of \eqref{eq:MPEC} is beyond the scope of this paper. 
However, in the next section, we introduce a BLP formulation and establish the relationship between the MPEC and the BLP. The BLP will provide an approach to find a globally optimal solution of \eqref{eq:MPEC} by solving only convex problems. We also present some key theoretical results regarding the optimal solutions in the following section.

\section{Theoretical Results}
\label{sec:Theo}

\subsection{Bilevel Reformulation}
The idea of reformulating the MPEC into a BLP is simple: to replace the lower-level market equilibrium conditions with a centralized optimization problem. This is based on the well-known first fundamental theorem in welfare economics, which states that a competitive equilibrium leads to a Pareto efficient market outcome. %(See, for example, Proposition 10.D.1 in \cite{Mas95}.) 
More specifically, let the set $\bm{X}$ denote the feasible region of the lower-level variables $\bm{x}$; that is, $\bm{X} = \{\bm{x}: \eqref{eq:nb-con1} - \eqref{eq:nb-con2}, \eqref{eq:iso_con1} - \eqref{eq:nb-price}\}$. We can define the following optimal value function:
\begin{equation}\label{eq:LowerNLP}
V(\tau^b, \tau^s) :=   \underset{\bm{x} \in \bm{X}}{max} \ \Pi\left[\bm{x}; (\tau^b, \tau^s)\right]. 
\end{equation}
Let $\bm{x}(\tau^b,\tau^s)$ denote the optimal solution of the above optimization problem with respect to a given $(\tau^b, \tau^s)$, which is a set-valued mapping in general. However, we state below that under mild conditions, the mapping is a singleton. 
\begin{lemma}\label{lem:unique}
With the inverse demand functions as in \eqref{demand} and \eqref{demandp} and under Assumption \ref{as:QuadraticCost}, for any given $(\tau^b, \tau^s)$ with $\tau^b \geq 0$ and $\tau^s \leq \tau^b$, an optimal solution of the optimization problem in \eqref{eq:LowerNLP} exists. In addition, the vectors $d, l, g$ and the quantity $z^s - z^b$ in such an optimal solution are all unique. %If $\tau^s < \tau^b$, then both $z^s$ and $z^b$ are also unique in an optimal solution. 
\end{lemma}
The proof is relatively straightforward, with the existence following from the coercieveness of the objective function over the feasible region $\bm{X}$, and the uniqueness as the result of strict convexity of the objective function.  Details are omitted here.  The complete BLP model is as follows. 
\begin{subequations}
\begin{align}
 \underset{\tilde{\tau}^b,\ \tilde{\tau}^s,\ \phi^{con},\ \phi^{pro}}{\mathrm{Minimize}} & 
 B\left[\tilde{\tau}^b, \tilde{\tau}^s, \phi^j_i; 
 d(\tilde{\tau}^b, \tilde{\tau}^s), z^b(\tilde{\tau}^b, \tilde{\tau}^s), p(\tilde{\tau}^b,\tilde{\tau}^s)\right] \label{eq:UpperObj} \\[-10pt] 
  \mathrm{subject\ to} \quad &  \sum_{i=1}^N \Big[-z_i^s(\tilde{\tau}^b, \tilde{\tau}^s)\tilde{\tau}^s  +  z_i^b(\tilde{\tau}^b, \tilde{\tau}^s)\tilde{\tau}^b + \nonumber \\
   &  \quad \quad d_i(\tilde{\tau}^b, \tilde{\tau}^s)\tau^b+\sum_{j \in {pro,con}} \phi^j_i \Big] = \Tau, \label{eq:BL_RevAdq} \\[-5pt]
  &  \tilde{\tau}^s \leq \tilde{\tau}^b,\ \tilde{\tau}^b,\ \phi^{con},\ \phi^{pro} \geq 0,\ \\
  %\tilde{\tau}^s \leq \tau^b, \\
  & (\tilde{\tau}^b, \tilde{\tau}^s) \in \underset{(\tau^b, \tau^s)}{\arg\max}\ V(\tau^b, \tau^s), \label{eq:OptimalTau} \\
  & (d_i(\tilde{\tau}^b, \tilde{\tau}^s),  z_i^s(\tilde{\tau}^b, \tilde{\tau}^s), z_i^b(\tilde{\tau}^b, \tilde{\tau}^s), 
  p(\tilde{\tau}^b, \tilde{\tau}^s)) \nonumber\\ 
  & \in  \bm{X}_{KKT}(\tilde{\tau}^b, \tilde{\tau}^s), \label{eq:UpperLastConstr}
\end{align}
\end{subequations}
where the set $\bm{X}_{KKT}(\tilde{\tau}^b, \tilde{\tau}^s)$ represents the KKT system of the lower-level optimization problem \eqref{eq:LowerNLP} with respect to $(\tilde{\tau}^b, \tilde{\tau}^s)$. 
Under the assumptions of Lemma \ref{lem:unique}, the lower-level optimization problem is clearly a convex optimization problem with all linear constraints. Hence, the KKT conditions are necessary and sufficient optimality conditions. By Lemma \ref{lem:unique}, with the uniqueness of optimal solutions corresponding to a given $(\tau^b, \tau^s)$, constraint \eqref{eq:BL_RevAdq} is well defined. However, the objective function \eqref{eq:UpperObj} may not be. %Although $z_i^b(\tilde{\tau}^b, \tilde{\tau}^s)$ may not be unique when $\tau^s = \tau^b$, by choosing $\tilde{z}_i^b := \max\{z^{b^*}_{i} - z^{s^*}_{i}, 0\}$ as discussed after Lemma \ref{lem:zszb}, the corresponding $\tilde{z}_i^b$ will be unique due to the uniqueness of the difference: $z^{b^*}_{i} - z^{s^*}_{i}$. 
The issue is $p(\tilde{\tau}^b, \tilde{\tau}^s)$, which is the Lagrangian multiplier (i.e., the LMP) associated with the flow balancing constraint \eqref{eq:nb-price} may not be unique, even when the primal variables are unique. If the set of multipliers is not a singleton, it is understood that the optimization problem chooses a $p(\tilde{\tau}^b, \tilde{\tau}^s)$ to minimize the objective function \eqref{eq:UpperObj}. 
%If the linear independence constraint qualification (LICQ) holds at the optimal solution with respect to $(\tilde{\tau}^b, \tilde{\tau}^s)$, then $p(\tilde{\tau}^b, \tilde{\tau}^s)$ is a single-valued mapping, and there will be no ambiguity with respect to the objective function in \eqref{eq:UpperObj}. 

To further ensure the validity of the BLP, we need to ensure the attainability of maximum of the optimal value function $V(\tau^b, \tau^s)$. To do so, we first argue that $(\tau^b, \tau^s)$ should be in a bounded polyhedral set. %Although $\tau^s$ is not explicitly bounded, if $\tau^s > \tau^b$, as discussed in Section \ref{subsec:Prosumer}, the prosumers' objecitve functions will be unbounded, and both $\bm{z^s}$ and $\bm{Z^b}$ will go to infinity. However, based on the definition of energy expenditure in \eqref{eq:inc_pro}, only the spending associated with $\bm{Z^b}$ appears. Hence, to minimize the energy equity index, $\bm{\phi}^{pro}$ needs to become $-\infty$ (meaning an unlimited subsidy from the government to consumers), or $\bm{\phi}^{con}$ has to be $+\infty$ (meaning consumers have to pay an infinite amount of fixed charge). Neither scenario is plausible. As such, we explicitly require $\tau^s \leq \tau^b$.
%we assume that both $\bm{\phi}^{pro}$ and $\bm{\phi}^{con}$ are in a bounded set $\Phi$ that they are large enough to still make the constraint \eqref{eq:BL_RevAdq} always satisfied. Yet, under this assumption, $t^s > t^b$ can never be an optimal solution; as otherwise, with $\bm{z^b}$ going to $+\infty$, the objective function \eqref{eq:UpperObj} will go to $+\infty$ with implicitly bounded  $\bm{\phi}^{pro}$ and $\bm{\phi}^{con}$. Any other feasible solution $(\tau^b, \tau^s, \bm{\phi}^{con}, \bm{\phi}^{pro})$ with $\tau^s \leq \tau^b$ will lead to a finite objective function value. 
We observe that, based on the KKT conditions \eqref{eq:d_KKT}, \eqref{eq:ProKKT_l}, and \eqref{eq:Pro_zb}, if $\tau^b > \max_{i=1, \ldots, N}\{P_i^0\}$, that is, if the volumetric charge is greater than the largest willingness-to-pay of any consumers/prosumers, then all the $d_i$'s and $l's$ will be 0 in an optimal solution, indicating no energy consumption at all due to the high costs. To avoid this, $\tau^b$ should be bounded above by $\max_{i=1, \ldots, N}\{P_i^0\}$, which we denote by $\widehat{\tau}^b$. Similarly, if $\tau^s$ is too negative, no prosumers would sell their self-produced energy, and all $z^s_i$ would be 0 in an optimal solution. Hence, we assume that $\tau^s$ is implicitly bounded below by $\underline{\tau^s} < 0$. Define the following set: 
\begin{equation} \label{eq:setT}
\mathcal{T} := \{(\tau^b, \tau^s): 0\leq \tau^b \leq \widehat{\tau}^b,\ \underline{\tau}^s \leq 
\tau^s \leq \tau^b\} \in \Re^2,    
\end{equation} which is a bounded polyhedron. 
For the optimal value function $V(\tau^s, \tau^b)$, we want to show that it is continuous over $\mathcal{T}$. 

%Note that the strict convexity assumption of the cost functions is sufficient for our purpose; the quadratic function form is just to simplify the proofs. 

Under Assumption \ref{as:QuadraticCost}, $V(\tau^b, \tau^s)$ is the optimal value function of a convex quadratic program with parameterization only on the linear term in the objective function. Properties of such functions are well established, and in this specific case, it is known that $V(\tau^b, \tau^s)$ is a continuous function. (See Theorem 47 in \cite{TerlakyOptValueFunc}.) Therefore, an optimal solution of 
$\max_{(\tau^b, \tau^s)\in \mathcal{T}} V(\tau^b, \tau^s)$ exists and is attainable, again by the Weierstrass Extreme Value Theorem. \vspace{-1em}
%consequently, the constraint \eqref{eq:OptimalTau} is well defined.     
%To avoid technical difficulties, we do not explicitly bound the variables $\bm{\phi}^{con}$ and $\bm{\phi}^{pro}$ (and they can be positive or negative, with a negative $\phi$ representing a subsidy from the government to consumers/prosumers); hence, constraint \eqref{eq:BilevelConstr} can always be satisfied. This being said, note that 
\subsection{Equivalence and Optimal Solutions}
Note that the relationship between the MPEC model and the BLP model is different than their usual relationship. In a typical BLP,  writing out the lower-level problem's KKT conditions will lead to an MPEC formulation. There, the upper-level decision maker does not optimize the lower-level objective function; that is, there is no optimization of the optimal value function $V(\tau^b, \tau^s)$ in \eqref{eq:OptimalTau}. 
It is needed here since the upper-level decision maker wants to maximize social surplus as well as maintaining equity among all energy consumers. 
%although the wholesale energy market is not under the state government's control, and hence, the bilevel formulation. 
Because of \eqref{eq:OptimalTau}, the BLP formulation \eqref{eq:UpperObj} -- \eqref{eq:UpperLastConstr} is not equivalent to the MPEC formulation \eqref{eq:MPEC} in general, due to the multi-objective decision-making in \eqref{eq:MPEC}; that is, there can be a solution that leads to a lower EEI-difference function $B$, but does not maximize the social surplus $\Pi$. 
%In this case, to convert to the MPEC formulation, the lower-level's objective function needs to appear in the upper-level's objective function as well, as in \eqref{eq:MPEC}. 
In the following, we show that if an optimal solution exists for each model, the MPEC and the BLP formulations are indeed equivalent when there are no additional constraints on the volumetric and fixed charges.  We then show that an optimal solution of the BLP does exist, with the optimal $\tau^{b^*} = \tau^{s^*} = 0$.   

%decides on the volumetric rates $\tau^b$ and $\tau^s$, and the fixed charges $\bm{\phi}^{con}$ and $\bm{\phi}^{pro}$. Then the wholesale market clears its supply and demand and returns an optimal solution $(\bm{d^*, z^{s^*}, z^{b^*}})$, along with the wholesale energy price $\bm{p^*}$,  so that the government agency can ensure equity by minimizing the energy expenditure incidence differences as in the objective function \eqref{eq:BilevelObj}, while ensuring that utilities' costs and spending can be covered by the constraint \eqref{eq:BilevelConstr}. 
 
%\textcolor{cyan}{Need to show that the MPEC and the bilevel problem have an optimal solution first}  

\begin{proposition}\label{pr:Equiv}
Assume that both the MPEC model \eqref{eq:MPEC} and the bilivel optimization model \eqref{eq:UpperObj} -- \eqref{eq:UpperLastConstr} have an optimal solution, which are denoted as $(\bm{x}^{MPEC}, \bm{\zeta}^{MPEC})$ and $\bm{\zeta}^{BLP}$, respectively. 
Under the assumptions in Lemma \ref{lem:unique}, 
and if the LICQ holds at $\bm{x}(\bm{\zeta}^{BLP})$ in the set $\bm{X}$, 
then the MPEC model \eqref{eq:MPEC} and the BLP model \eqref{eq:UpperObj} -- 
\eqref{eq:UpperLastConstr} are equivalent.
\end{proposition}
Due to the page limit, we can only show the sketch of the proof. The keys are: (i) the centralized optimization problem of $\max_{\bm{x} \in \bm{X}}\Pi\left[\bm{x}; (\tau^b, \tau^s)\right]$ and the KKT systems (\ref{eq:d_KKT}), (\ref{eq:ProKKT_l})-(\ref{eq:ProKKT_delta}),  (\ref{eq:foc-iso-x})-(\ref{eq:foc-iso-g}) are equivalent, under the conditions in this proposition; and (ii) with an arbitrary $(\tau^b, \tau^s) \in \mathcal{T}$, the EEI-difference function $B$ can be minimized to 0 by equating the EEI of conventional consumers and prosumers, which, together with the constraint \eqref{eq:nb}, leads to the following linear system of equations with respect to $(\phi^{con}, \phi^{pro})$: 
\begin{align}
	\frac{(p_i + \tau^b)d_i+\phi^{con}_i}{\textcolor{black}{I^{con}_i}} = \frac{(p_i + \tau^b)z^b_i+ \phi^{pro}_i+C_i(g_i) + SC_i}{\textcolor{black}{I^{pro}_i}},  \nonumber\\ 
	\mathrm{for\ all\ } i = 1, \ldots, N  \nonumber\\
\sum_{i=1}^N \Bigg(\textcolor{black}{-z_i^s\tau^s}  +  z_i^b\tau^b +d_i\tau^b +\sum_{j \in {pro,con}} \phi^j_i \Bigg)= \Tau. \label{eq:EqualIncidence}
\end{align} 
The above system has $N + 1$ equations but $2N$ variables. The coefficient matrix clearly has a full-row rank. Hence, for any given $(\tau^b, \tau^s)$, a solution of $(\phi^{con}, \phi^{pro})$ always exists.

Proposition \ref{pr:Equiv} is established under the assumption that an optimal solution exists for both the MPEC and the BLP. In the following, we show that the BLP indeed has an optimal solution and in such an optimal solution,  $\tau^{b^*} = \tau^{s^*} = 0$. This solution is trivially optimal for the MPEC model as well. 
%\hl{The MPEC model has an optimal solution can use the same arguments in Lemma Unique and coerciveness/Weierstrass Theorem.}   

%\begin{assumption}\label{as:LessDER}
%We assume that $
%\sum_{i=1}^N K_i \leq \sum_{i=1}^N Q_i^0. 
%$
%\end{assumption}

\begin{proposition}\label{pr:OptimalTau}
Under the assumptions of Lemma \ref{lem:unique} and Assumption \ref{as:QuadraticCost}, we have 
$\max_{(\tau^b, \tau^s) \in \mathcal{T}} V(\tau^b, \tau^s) = V(0, 0)$. 
\end{proposition}
\begin{proof} See appendix. 
\end{proof} 
\begin{remark}
With $\tau^{b^*} = \tau^{s^*} = 0$, as argued in the proof of Proposition \ref{pr:Equiv}, the system of equations \eqref{eq:EqualIncidence} always has a solution, which will make the $B$ function equal to zero; hence, they together form an optimal solution of the BLP formulation, which can be easily seen to be a globally optimal solution of the MPEC formulation. Obtaining such a solution can be done in two steps: first, solve the lower-level market equilibrium (either by solving a complementarity problem or an optimization problem as in \eqref{eq:LowerNLP}, with $(\tau^b,\tau^s) = (0,0)$; second, solve the linear system equations \eqref{eq:EqualIncidence}. Both steps can now be done with efficient algorithms since no non-convex problems are involved. 
One thing to note is that since the system equations have fewer equations than the variables when $N > 1$, the solution $\phi$'s are not unique. In this case, we can minimize $||\phi||_2^2$ over all solutions, which is a convex quadratic program and can be solved efficiently. 
\end{remark}

\begin{remark}
Proposition \ref{pr:OptimalTau} is consistent with a well-known fact in economics: under perfect competition, \emph{laissez-faire} is the first-best equilibrium. Any type of tax or subsidy can only reduce economic efficiency and social surplus.  A fixed-charge-only tariff, however, would likely to encourage excessive energy use, as customers' energy bills are not tied to energy usage at all. Hence, in reality, almost all tariffs consist of both volumetric and fixed charges. In Section \ref{sec:case}, we numerically study the impact of different levels of volumetric charges on market outcomes and energy equity. However, one thing to note is that with additional requirements (aka constraints)on the fixed and volumetric charges, 
%such as a certain percentage of a utility's fixed cost $\Tau$ having to be recovered from volumetric charges, 
the system of equations \eqref{eq:EqualIncidence} (together with the additional constraints) may no longer have a solution, which means that the EEI difference function can no longer be zero. (This is indeed what we observe in certain numeric cases.) In this case, the equivalence between the MPEC and the BLP model may also break down.  \vspace{-1.2em}
%First-best versus second-best (other constraints to restrict $\tau$ and $\phi$. 
\end{remark}

\subsection{Model and Theoretical Results under Stochasticity}
The above analyses are performed based on deterministic models. 
%In reality, many parameters are stochastic, including the DER output $R_i$, the generation capacity $G_h$, the transmission line capacity $T_k$, and fuel costs $C_h$, among others. 
In this section, we show that the main result in Proposition \ref{pr:OptimalTau} still holds under uncertainties, regardless of the probability distributions. To do so, however, we need to present a formulation under stochasticity. 
Let 
%$\omega \in \Omega \subset \Re^w$ represent a random event, with $\Omega$ being the sample space and $w$ being a general dimension. We use the random vector 
$\bm{\xi} \in\Re^w$ denote the collection of all the random variables with a generic dimension of $w$ (such as prosumers' solar output $R_i$, generation capacity $G_h$, etc). It is defined in the probability space $(\Omega, \mathscr{A}, \mathcal{P})$, with $\Omega$ being the sample space of all the uncertainties, $\mathscr{A}$ the sigma field of $\Omega$ and $\mathcal{P}$ a probability measure in $(\Omega, \mathscr{A})$; that is, $\bm{\xi}$ is a measurable mapping from $\Omega$ to a set $\Xi \in \Re^w$.
Let $\bm{x}^{\bm{\xi}}$ and $\bm{X}(\bm{\xi})$ respectively denote the lower-level variables and feasible region corresponding to a realization of the random vector $\bm{\xi}$. Furthermore, let $\Pi[\bm{x}^{\bm{\xi}}; (\tau^b, \tau^s), \bm{\xi}]$ represent the lower-level objective function under uncertainty, with a given volumetric rate $(\tau^b, \tau^s)$. Then we can define the expected optimal value function of the lower-level problem as follows:
\begin{equation}\label{eq:EV}
EV(\tau^b, \tau^s) :=  \mathbb{E}_{\bm{\xi}} 
\left\{\underset{\bm{x}^{\bm{\xi}} \in \bm{X}(\bm{\xi})}{max}\Pi\left[\bm{x}^{\bm{\xi}}; (\tau^b, \tau^s), \bm{\xi}\right]\right\}.
\end{equation}
To ensure that the above expectation is well-defined, we need the following assumption. 
\begin{assumption}\label{as:FiniteMoment}
All the random functions in $\Pi\left[\bm{x}^{\bm{\xi}}; (\tau^b, \tau^s), \bm{\xi}\right]$ and $\bm{X}(\bm{\xi})$ have finite moments. 
\end{assumption}

Under Assumption \ref{as:FiniteMoment}, for any $(\tau^b, \tau^s)\in\mathcal{T}$, with $\mathcal{T}$ defined in \eqref{eq:setT}, the expectation in \eqref{eq:EV} is well defined following the same argument as in \cite{PangSen}. 
%since the feasible region $X(\bm{\xi})$ is always feasible (the zero vector is always in $X(\bm{\xi})$ for any $\bm{\xi}\in \Xi$), the optimal solutions are unique with a given $(\tau^b, \tau^s)$ and $\bm{\xi}\in\Xi$, per Lemma \ref{lem:unique}, and the set $\mathcal{T}$ is compact. 
With a finite expected value in \eqref{eq:EV}, since expectation preserves convexity (see, for example, Sec. 3.2.1 in \cite{Boyd_ConvexOpt}), we have that $EV(\tau^b, \tau^s)$ is also a convex function with respect to $(\tau^b, \tau^s) \in \mathcal{T}$. Hence, 
the extension of Proposition \ref{pr:OptimalTau} to the stochastic case is straightforward, and we only state the result below, omitting the proof. 
\begin{proposition}\label{pr:SOptimalTau}
Under the assumptions of Lemma \ref{lem:unique}, and Assumption \ref{as:QuadraticCost} and \ref{as:FiniteMoment}, we have 
$\max_{(\tau^b, \tau^s) \in \mathcal{T}} EV(\tau^b, \tau^s) = EV(0, 0)$.  
\hfill$\Box$
\end{proposition} 
With the lower-level optimal value function defined, we can write out the stochastic version of the upper-level problem now. Note that the upper-level decisions on the retail rates have to be made before lower-level uncertainties are realized; that is, they are the here-and-now type of decisions.
Since the lower-level optimal solutions $d$, $z^b$, and $p$ depend on the random vector $\bm{\xi}$, 
%of which we do not specify a specific (joint) probability distribution, 
if we simply require that the revenue adequacy constraint \eqref{eq:nb} is held for all $\bm{\xi}$ (almost surely), it will likely to be always infeasible. To remedy this, a natural idea is to make  constraint \eqref{eq:nb} a chance constraint as follows:
\begin{align}
 \underset{\tilde{\tau},\ \phi}{min}\ & \mathbb{E}_{\bm{\xi}}\left[B\left(d_i^{\bm{\xi}}(\tilde{\tau}^b, \tilde{\tau}^s), \tilde{\tau}^b, \phi^j_i\right)\right]  \label{eq:SBilevelObj}\\
 %+ M||\bm{\zeta}^+||_2^2\label{eq:SBilevelObj}\\
   \mathrm{s.t.} \quad \quad \ &  \mathcal{P}\left\{\sum_i \Big[-z_i^{s^{\bm{\xi}}}(\tilde{\tau}^b, \tilde{\tau}^s)\tilde{\tau}^s  + z_i^{b^{\bm{\xi}}}(\tilde{\tau}^b, \tilde{\tau}^s)\tilde{\tau}^b + \right.\nonumber \\ 
   & \left. \quad  d_i^{\bm{\xi}}(\tilde{\tau}^b, \tilde{\tau}^s)\tau^b+\sum_{j \in {pro,con}} \phi^j_i \Big] \geq \Tau \right\} \geq 1 - \epsilon \nonumber \\
   %\label{eq:SBilevelConstr} \\
  & (\tilde{\tau}^s, \tilde{\tau}^b) \in \underset{(\tau^s, \tau^b)}{\arg\max}\ EV(\tau^s, \tau^b), \nonumber
\end{align}
where $\epsilon$ is a pre-specified parameter. 
Solving the chance-constrained stochastic program (SP) \eqref{eq:SBilevelObj} is much more involved than its deterministic counterpart, even when Proposition 
\ref{pr:SOptimalTau} holds. In this case, sample average approximation (SAA) methods developed for chance-constrained SPs, such as \cite{SAA_CC}, can be directly applicable here. However, if there are other requirements on the volumetric charges such that they cannot be zero, then specialized (and likely iterative) algorithms need to be developed to solve \eqref{eq:SBilevelObj}, since it also includes an optimal value function in its constraints. Developing such algorithms (and presenting numerical results under stochasticity) is deferred to future research.  \vspace{-0.5em}

\section{Numerical Case Study \vspace{-0.5em}} \label{sec:case}
\subsection{Setup}
To illustrate the effects of optimal pricing schemes, we apply the models developed in Section \ref{sec:mod} to a case study considering a three-node network with three firms, ten generating units, and three transmission lines.
This setup is sufficiently general because it allows firms 
%to own facilities and 
compete across different locations subject to transmission constraints. %\footnote{%A similar setup was previously applied to examining carbon leakages under California climate change policy \cite{chen11}. 
The three-node network is the simplest that allows for looped flows, which is important in modeling a power grid.  
%Our intention is to analyze the impacts of different pricing policies on the energy expenditure incidence.
%Thus, we believe that this setup is reasonable and sufficient for our purposes.
%}
We assume that a daily fixed cost of \$80k  to be reimbursed to a utility. %based on a volumetric tariff. 
%For prosumers' DER capacity $R$, we consider for cases of $R=25$MW and $R=150$MW, which are carefully chosen with one corresponding to insufficient DER generation for prosumers,  while the other with excess DER generation from prosumers. 
%\textcolor{black}{We then enumerate over four scenarios for $R=25$MW and $R=150$MW assuming that a regulatory scheme first fixes fraction of cost recovery by a volumetric charge.
%Note that these two levels of $R$ are carefully chosen with one corresponding to insufficient DER generation for prosumers,  while the other with excess DER generation from prosumers.  } 

%\hl{For a meaningful comparison, we first establish the baseline by solving a cost-minimizing problem subject to fixed demand, assuming a demand elasticity equal to 0.05 to recover $P^0_i$ and $Q^0_i$  for all nodes $i$. (AL: I still don't understand this sentence.)}
Consumers are grouped into three income levels: high, medium and low, residing at nodes A, B, and C, respectively.
%\footnote{While there is income heterogeneity among consumers who live in the same state, state-level median incomes and average power prices are highly correlated as consumers who cannot afford high energy prices (and other expenses) would move somewhere more affordable. \hl{(AL: I don't quite get it what point this footnote is trying to make. Can we delete it?)}} 
The baseline daily demand of the low-income group is assumed to be 20 kWh.
The daily demand of the medium- and high-income groups are assumed to be 25\% and 50\% larger than the low-income group, broadly consistent with the data from the 2015 RECS survey \cite{recs15}.
Given the assumed fixed demand in each node, we then recover the number of households in each income group.
(Proportion of households among income groups are also compatible with the 2015 RECS results.) 
The income level is obtained by assuming that electricity expenditure is 1.5\% of the income in each group.\footnote{This represents the ``energy equity'' case at the baseline. 
The 1.5\% is at the lower end based on 2015 RECS. 
However, our interest lies on the relative changes of the energy incidence when the power sales by prosumers are subject to different tariff designs.
This assumption is not essential.}  
Finally, based on RECS 2015, we assume that 20\% of the households in high-income group (or 3,067 households) own rooftop solar energy with a capacity of 8 KW each household or 25 MW in total. 
Thus, in the extreme case during a sunny summer day, we consider a daily solar output from prosumers to be $R = 150$ MWh; while in a cloudy/raining winter day, it generates only $R=25$ MWh. The numbers are carefully chosen with one corresponding to insufficient DER generation for prosumers,  while the other with excess DER generation from prosumers.   
%\hl{YC: The following sentence needs your help to fill out per our discussion while you were on vacation...}
For the sunk cost $SC$ to be included in prosumers' energy expenditure in \eqref{eq:inc_pro}, we assume a daily cost \$5/day.\footnote{This is calculated approximately based on \$30K initial investment on solar panels (including installation) and an assumed break-even period of 15 years. This is also in the ballpark of rental costs of solar panels.}
%\hl{(AL: connection needs to be made between the solar capacity and the value $R$. Should $R$ be measured in MW, MWh, or MW/hour?)}
%\hl{YC: R is in MWh. The total of number of prosumers household is 3,067. With 8kw of each household, it is about 25 MW. If operating for 8 hours, in the extreme case it can produce 150MWh/d. }
A 25 MW backup generator or energy storage is assumed for the prosumers. 
The MPEC formulation is used for computation, and it is written in AMPL and solved by Knitro solver version 12.4 on a Mac Book Pro with 2.8 GHz Quad-Core and Intel Core i7. \vspace{-1em}
%\hl{How is the problem solved, what solver, which version, on what computer?}

\subsection{Main Results}
First, we add additional constraints to the upper-level problem and require that a certain percentage of utility's fixed cost $\Tau$ to be recovered through volumetric charge. More specifically, we consider two cases: 10\% and 90\% of $\Tau$ to be recovered by volumetric charges. Later we will compare such results with the case without the additional constraints (i.e., the original formulation as in \eqref{eq:MPEC}). 
The results of the volumetric charge $= 10\% \Tau$ and $90\%\Tau$ for the cases with renewable output equal to 25 MWh and 150 MWh are reported in Tables \ref{tab:25} and \ref{tab:150}, respectively.
%Each table includes three parts related to prosumers, the wholesale market, and the distribution of economic rent in the upper, middle, and lower panels, respectively.

Table \ref{tab:25} indicates a significant increase in volumetric charge when a higher percentage of $\Tau$ is required to be recovered from the actual energy use. Specifically,  $\tau^b$ increases from \$5.25/MWh under $10\%\Tau$  to \$49.18/MWh under $90\%\Tau$.   
%The retail rate facing prosumers at node A increases from \$83.2/MWh (=77.93+5.27) to \$126.84/MWh (=77.66+49.18) under 10\% and 90\%, respectively. 
In the case of 25 MWh of DER output, prosumers in both cases are in a net-buying position.
%Under 10\%, they purchase 45.68 MWh together with 25MWh to meet their load of 95.68MWh. 
%Under 90\%, facing a high retail rate, they purchase less (43.10 MWh) and consume less (93.10 MWh).
The prosumers benefit from the case of requiring a higher percentage of volumetric charge, with an increase of surplus from \$73.47K to \$78.23K and a decline of energy expenditure incidence from 1.34\% to 1.29\%. 
Note that prosumers' energy incidence is less than that of consumers under the $90\%\Tau$ case; meaning that with this requirement, the system of equations \eqref{eq:EqualIncidence} no longer has a solution of $(\phi^{con}, \phi^{pro})$ that can lead to zero value of the EEI difference function $B$. This indicates that energy equity can no longer be maintained across different income groups.  
%That is, \$8,000 (or 10\% of $\Tau$ in (9)) under the 90\% is not adequate for the PUC to maintain energy equity across different income groups.    
The generation from the wholesale market is 1,519.17 MWh and 1,464.09 MWh, under the 10\%- and 90\%-volumetric charge requirement, respectively. This is so because the higher energy prices faced by consumers in the $90\%\Tau$ case suppress demand, including that of prosumers, by a margin of 3.6\%. 

%In general, consumers benefit from cases allocating more to volumetric charge.
%This is mainly because demand response acts as a means to soften the blow of recovering fixed costs. 
%On the other hand, producers surplus declines significantly due to lower wholesale prices induced by high retail rates resulting from a large volumetric tariff. 
%Overall, the declines in producers' surplus is not adequately recouped by the increase in consumers surplus, leading to a decline in the total surplus of \$6.09K (=\$819.60K-\$813.51K).  
%However, the increase in prosumers' surplus under the 10\% is more than offset the decline in the wholesale surplus, leading to larger total social surplus.

\begin{table}[!ht]
\renewcommand{\arraystretch}{0.9}
\caption{Results of 25 MWh Renewable case: 10\%  vs. 90\% volumetric charge}
\resizebox{\columnwidth}{!}{
\label{tab:pc} 
\centering
\begin{tabular}{l|rrrrrr}	 \hline
Variables\textbackslash Volumetric charge           & \multicolumn{3}{r}{10\%} & \multicolumn{3}{r}{90\%}     \\\hline						
%\color{black}{Renewable output} [MW] & & 20 & 160\\  \hline
Volumetric charge [\$/MWh] & \multicolumn{3}{r}{5.27}  & \multicolumn{3}{r}{49.18} \\
Prosumer's sale(+)/buy(-)  [MWh] & \multicolumn{3}{r}{-45.68 } & \multicolumn{3}{r}{-43.10}  \\ 
%Prosumer's purchase [MW]            & 66.27  & 44.11   & 0.00           \\ 
Prosumer's load [MWh]        & \multicolumn{3}{r}{95.68}  & \multicolumn{3}{r}{93.10}  \\ % due to backup generation that incite the prosumers to consume more. 	
Backup generation [MWh]     &  \multicolumn{3}{r}{25.0}     & \multicolumn{3}{r}{25.0} \\ 
Prosumer  surplus [\$K] 		 & \multicolumn{3}{r}{73.47}  & \multicolumn{3}{r}{78.23}        \\
Prosumer  incidence [\%] 		 & \multicolumn{3}{r}{1.34}  & \multicolumn{3}{r}{1.29}\\
Prosumer fixed charge [\$/household]     &  \multicolumn{3}{r}{2.18}     & 
\multicolumn{3}{r}{0.00031} \\\hline
  Variables\textbackslash Nodes &  A  & B & C & A & B & C    \\ \hline
Conventional demand [MWh]  &  382.70 & 592.71 & 498.08  & 372.38 & 577.39 & 469.46    \\
Fixed Charge [\$/household] & 1.29 & 1.25& 0.93&0.00 &0.35 & 0.01\\
Power price [\$/MWh]  &   77.93  & 52.63  & 37.84 & 77.66  & 40.20 & 37.50          \\
Energy incidence [\%] & 1.34 & 1.34 & 1.34 & 1.32 & 1.27 &1.28\\
%Sales-weighted power price [\$/MWh]    &		31.03  & 30.89  \\ \hline	
%Consumer surplus [\$K] & 285.5  & 320.9 & 170.4 & 285.5  & 321.0 & 170.4  \\ % could this be 
Conventional generation [MWh] & \multicolumn{3}{r}{1,519.17} & \multicolumn{3}{r}{1,464.09} \\ \hline
Total consumer surplus [\$K] & \multicolumn{3}{r}{806.51} & \multicolumn{3}{r}{807.80} \\
Producer surplus [\$K] & \multicolumn{3}{r}{11.51} & \multicolumn{3}{r}{4.38} \\
ISO's revenue [\$K] & \multicolumn{3}{r}{1.58} & \multicolumn{3}{r}{1.33} \\
Wholesale surplus [\$K] & \multicolumn{3}{r}{819.60} & \multicolumn{3}{r}{813.51} \\  \hline
Total social surplus [\$K]& \multicolumn{3}{r}{893.07} & \multicolumn{3}{r}{891.74} \\ \hline
\end{tabular}}
\label{tab:25}
\end{table}

We now turn to discuss Table \ref{tab:150}, the case when the DER output equals 150 MWh. In this case, prosumers have excess output and are in a net-selling position, while subject to $\tau^s$ when selling to the grid. 
%DER generation and sell surplus energy into the grid. 
%Different from Table \ref{tab:25}, now, prosumers sell power into the grid and are subject to $\tau^s$, which is shown in the parenthesis in ``Volumetric Charge'' row.   
The fact that $\tau^s <0$ (-\$3.99/MWh and -\$13.97/MWh under the $10\%\Tau$ and $90\%\Tau$ cases, resp.) indicates that prosumers should contribute to recovering fixed costs when selling power to the grid (i.e., net billing with sales payment less than the utility retail rate is more optimal than net metering). 
This is consistent with the recommendation by the CPUC concerning the recent debate to revamp the net-metering policies in California \cite{borenstein22}.
%Moreover, prosumers sell less into the grid even just marginally as they  earn significantly less per MWh, i.e., \$72.65 (=76.64-3.99) under 10\% and 62.5 (76.47-13.97) under 90\% in Table \ref{tab:150}.
%However,  prosumers can effectively avoid fixed cost under the 90\% case, leading to higher surplus of \$89.93K compared to \$79.57K under 10\%.   
Overall, the same observations about the surplus distribution as in Table \ref{tab:25} emerge in Table \ref{tab:150}. 
More importantly, the $90\%\Tau$ requirement leaves the PUC with no adequate fund to maintain energy equity, leading to a divergence of energy incidence across different income groups.   \vspace{-1em}
% \cite{chen21}.

\begin{table}[!ht]
\renewcommand{\arraystretch}{0.9}
\caption{Results of 150 MWh Renewable case: 10\%  vs. 90\% volumetric charge}
\resizebox{\columnwidth}{!}{
\label{tab:pc} 
\centering
\begin{tabular}{l|rrrrrr}	 \hline
Variables\textbackslash Volumetric Charges           & \multicolumn{3}{r}{10\%} & \multicolumn{3}{r}{90\%}     \\\hline						
%\color{black}{Renewable output} [MW] & & 20 & 160\\  \hline
Volumetric charge [\$/MWh] & \multicolumn{3}{r}{ 5.21 (-3.99)}  & \multicolumn{3}{r}{\textcolor{black}{49.93 (-13.97)}} \\
Prosumer's sale(+)/buy(-)  [MWh] & \multicolumn{3}{r}{ 78.70} & \multicolumn{3}{r}{78.10}  \\ 
%Prosumer's purchase [MW]            & 66.27  & 44.11   & 0.00           \\ 
Prosumer's load [MWh]        & \multicolumn{3}{r}{96.30}  & \multicolumn{3}{r}{\textcolor{black}{96.90}}  \\ % due to backup generation that incite the prosumers to consume more. 	
Backup generation [MWh]     &  \multicolumn{3}{r}{25.0}     & 
\multicolumn{3}{r}{25.0} \\ 
Prosumer  surplus [\$K] 		 & \multicolumn{3}{r}{79.57}  & 
\multicolumn{3}{r}{\textcolor{black}{80.93}}        \\	
Prosumer incidence [\%]     &  \multicolumn{3}{r}{1.30}     & 
\multicolumn{3}{r}{1.06} \\ 
Prosumer fixed charge [\$/household]     &  \multicolumn{3}{r}{3.31}     & 
\multicolumn{3}{r}{2.61} \\
\hline
Variables\textbackslash Nodes &  A  & B & C & A & B & C    \\ \hline
Conventional demand [MWh]  &  383.02  & 593.02 & 498.11 & \textcolor{black}{372.49}  & \textcolor{black}{577.02} & \textcolor{black}{470.76}    \\
Fixed Charge [\$/household] & 1.12 & 1.19& 0.88&0.00 &0.00 & 0.00\\
Power price [\$/MWh]  &   76.64  & 52.03 & 37.85  & \textcolor{black}{76.47}  & 40.20 & 37.49          \\	
Energy incidence [\%] & 1.30 & 1.30 & 1.30 & 1.32 & 1.11 &1.28\\
Conventional generation [MWh] & \multicolumn{3}{r}{1,395.48} & \multicolumn{3}{r}{\textcolor{black}{ 1,342.17}} \\ \hline
Total consumer surplus [\$K] & \multicolumn{3}{r}{810.90} & \multicolumn{3}{r}{\textcolor{black}{815.18}} \\
Producer surplus [\$K] & \multicolumn{3}{r}{10.73} & \multicolumn{3}{r}{3.99} \\
ISO's revenue [\$K] & \multicolumn{3}{r}{1.52} & \multicolumn{3}{r}{1.3} \\
Wholesale surplus [\$K] & \multicolumn{3}{r}{823.15} & \multicolumn{3}{r}{\textcolor{black}{820.47}} \\  \hline
Total social surplus [\$K]& \multicolumn{3}{r}{ 902.72} & \multicolumn{3}{r}{\textcolor{black}{901.41}} \\ \hline
\end{tabular}}
\label{tab:150}
\end{table}

For comparison purposes, Table \ref{tab0} provides the results without any requirements on volumetric or fixed charges. 
As proven in Proposition \ref{pr:OptimalTau}, the optimal $\tau^{b^*} = \tau^{s^*} = 0$, and the total social surplus in the cases of $R= 25$ MWh and $R = 150$ MWh are higher than their counterparts in Tables \ref{tab:25} and \ref{tab:150}, consistent with the theoretical results in Section \ref{sec:Theo}. 

%Overall, we find that the total social surplus under the $R= 25$ MWh and $R = 150$ MWh cases is higher than their counterparts in Tables \ref{tab:25} and \ref{tab:150}. 
%For example, the total surplus of 150 MWh equals \$902.73k in Table \ref{tab0}, which is larger than either \$902.72k and \$901.41k reported in Table \ref{tab:150} under the 10\% and 90\% cases.
%The numerical results are consistent with theoretical results in Section \ref{sec:Theo}.

\begin{table}[!ht]
\renewcommand{\arraystretch}{0.9}
\caption{Results of 25 and 150 MWh Renewable case under 0\% volumetric charge}
\resizebox{\columnwidth}{!}{
\label{tab:pc} 
\centering
\begin{tabular}{l|rrrrrr}	 \hline
Variables\textbackslash Renewable            & \multicolumn{3}{r}{25 MWh} & \multicolumn{3}{r}{150 MWh}     \\\hline						
%\color{black}{Renewable output} [MW] & & 20 & 160\\  \hline
Volumetric charge [\$/MWh] & \multicolumn{3}{r}{0 (0)}  & \multicolumn{3}{r}{0 (0)} \\
Prosumer's sale(+)/buy(-)  [MWh] & \multicolumn{3}{r}{-45.97 } & \multicolumn{3}{r}{78.94}  \\ 
%Prosumer's purchase [MW]            & 66.27  & 44.11   & 0.00           \\ 
Prosumer's load [MWh]        & \multicolumn{3}{r}{95.97}  & \multicolumn{3}{r}{96.06}  \\ % due to backup generation that incite the prosumers to consume more. 	
Backup generation [MWh]     &  \multicolumn{3}{r}{25.0}     & \multicolumn{3}{r}{25.0} \\ 
Prosumer  surplus [\$K] 		 & \multicolumn{3}{r}{73.24}  & \multicolumn{3}{r}{79.62}        \\
Prosumer incidence [\%]     &  \multicolumn{3}{r}{1.36}     & 
\multicolumn{3}{r}{1.33} \\ 
Prosumer fixed charge [\$/household]     &  \multicolumn{3}{r}{2.33}     & 
\multicolumn{3}{r}{3.40} \\\hline
  Variables\textbackslash Nodes &  A  & B & C & A & B & C    \\ \hline
Conventional demand [MWh]  &  383.94 & 592.70 & 501.30  & 384.25 & 593.02 & 501.31    \\
Fixed Charge [\$/household] & 1.52 & 1.31& 1.08& 1.46 & 1.25 & 1.03\\
Power price [\$/MWh]  &   77.95  & 57.91  & 37.88 & 76.64  & 57.26 & 37.88          \\
Energy incidence [\%] & 1.36 & 1.36 & 1.36 & 1.33 & 1.33 &1.33\\
%Sales-weighted power price [\$/MWh]    &		31.03  & 30.89  \\ \hline	
%Consumer surplus [\$K] & 285.5  & 320.9 & 170.4 & 285.5  & 321.0 & 170.4  \\ % could this be 
Conventional generation [MWh] & \multicolumn{3}{r}{1,523.94} & \multicolumn{3}{r}{1,399.64} \\ \hline
Total consumer surplus [\$K] & \multicolumn{3}{r}{803.58} & \multicolumn{3}{r}{807.73} \\
Producer surplus [\$K] & \multicolumn{3}{r}{14.45} & \multicolumn{3}{r}{13.63} \\
ISO's revenue [\$K] & \multicolumn{3}{r}{1.80} & \multicolumn{3}{r}{1.74} \\
Wholesale surplus [\$K] & \multicolumn{3}{r}{819.83} & \multicolumn{3}{r}{823.11} \\  \hline
Total social surplus [\$K]& \multicolumn{3}{r}{893.08} & \multicolumn{3}{r}{902.73} \\ \hline
\end{tabular}}
\label{tab0}
\end{table}

\begin{figure}[htbp] %  figure placement: here, top, bottom, or page
\centering
\includegraphics[width=3.45in]{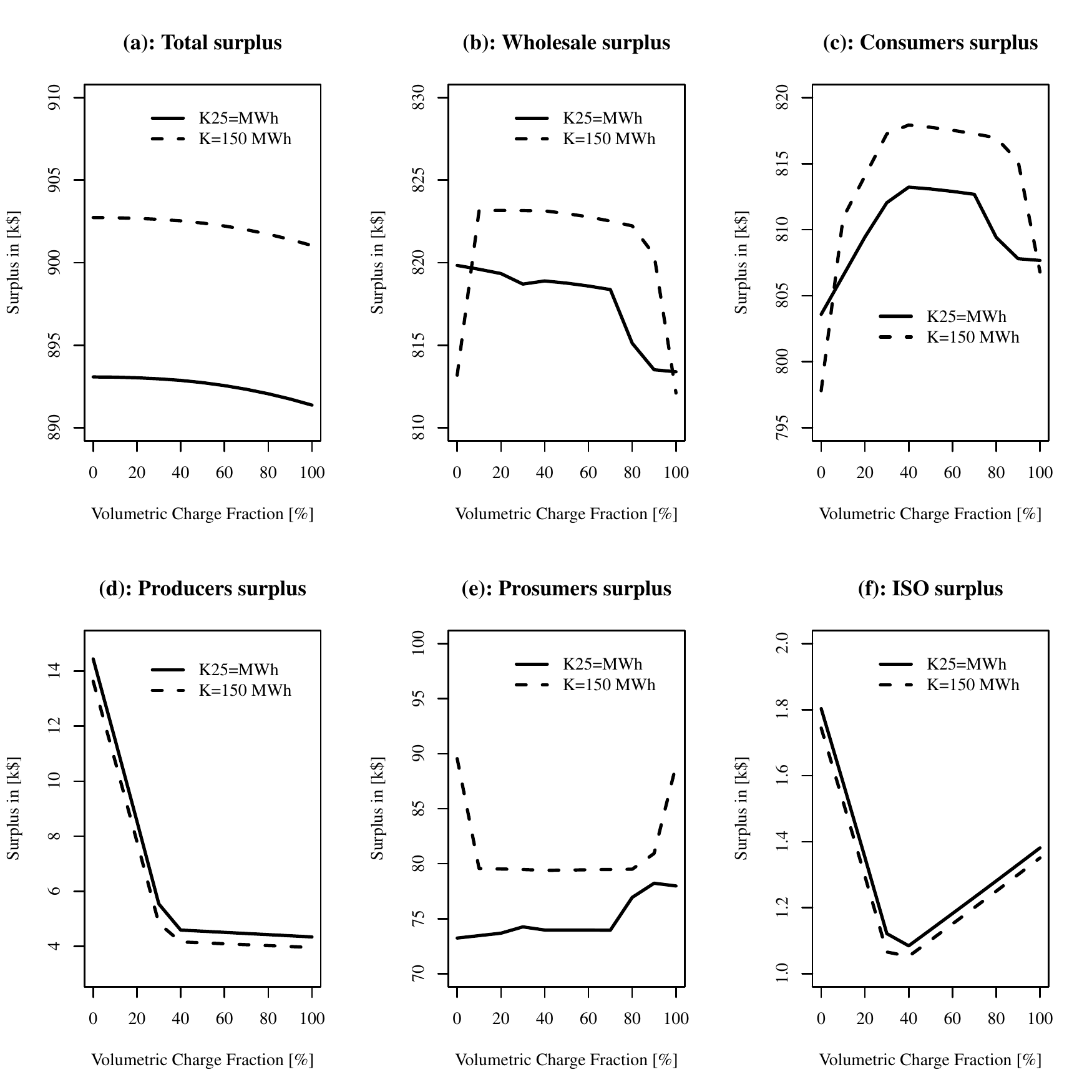}\vspace{-1em}
\caption{Distribution of surplus}
\label{fig:2}
\end{figure}

\textcolor{black}{Figure \ref{fig:2} plots the rent distribution among various entities in the market against the fraction of fixed costs assigned to the volumetric rate.
The palpable difference of prosumer's surplus in Figure \ref{fig:2}(e) of R=25 MWh cf. R=150 MWh cases affects the wholesale market surplus in Figure \ref{fig:2}(b) and deserves some explanation.
When R=25 MWh, the prosumers do not have enough DER generation and need to buy energy from the grid. 
An increase in the volumetric charge in x-axis provides an opportunity for the prosumer to ``avoid'' fixed charge via increasing self reliance or reducing consumption.
As a result, its surplus steadily increases until the volumetric charge becomes 100\%, where $\tau^b=\$60$/MWh or nearly 50\% of the retail price at node A, 
%(see Figure \ref{fig:3}, 
an unbearably high level.}

\textcolor{black}{On the contrary, prosumers sell energy to the grid when R=150MWh.
Its surplus in this situation is affected by two counteracting forces. 
On the one hand, its increased contribution to fixed cost through $\tau^s$ reduces its surplus until $\tau^s$ drops to approximately -\$10/MWh; beyond which its surplus is adequately offset by a decline in fixed  charge (or an increase in the volumetric charge fraction), leading its surplus to be leveled around \$80k until the volumetric charge fraction is greater than $90\%\Tau$. 
When the volumetric charge fraction is 100\%, the prosumer can avoid the fixed cost entirely, leading to a surge in its surplus as shown in Figure \ref{fig:2}(e).}
\textcolor{black}{With regard to consumers, their surplus is affected by retailed power prices and allotted fixed costs.
As alluded to earlier, forgo consumption is one way to minimize the impact of fixed cost recovery.
This strategy effectively enhances consumers' surplus until the allotted fraction of volumetric charge equal to 40\%.
However, beyond this level, the retail power prices become too high, leading to a drop in the surplus.}

\textcolor{black}{Overall, we observe that the wholesale surplus in Figure \ref{fig:2}(b) continues to decline with the increase in the volumetric fraction under R=25MWh, forming a concave curve due to the impacts on consumer surplus in \ref{fig:2}(c). Finally, the changes in consumers surplus is ``neutralized'' by the changes of prosumers' surplus in Figure \ref{fig:2}(e) under R=150MWh, leading to the total surplus in Figure \ref{fig:2}(a).} 

\begin{figure}[htbp] %  figure placement: here, top, bottom, or page
\centering
\includegraphics[width=3.45in]{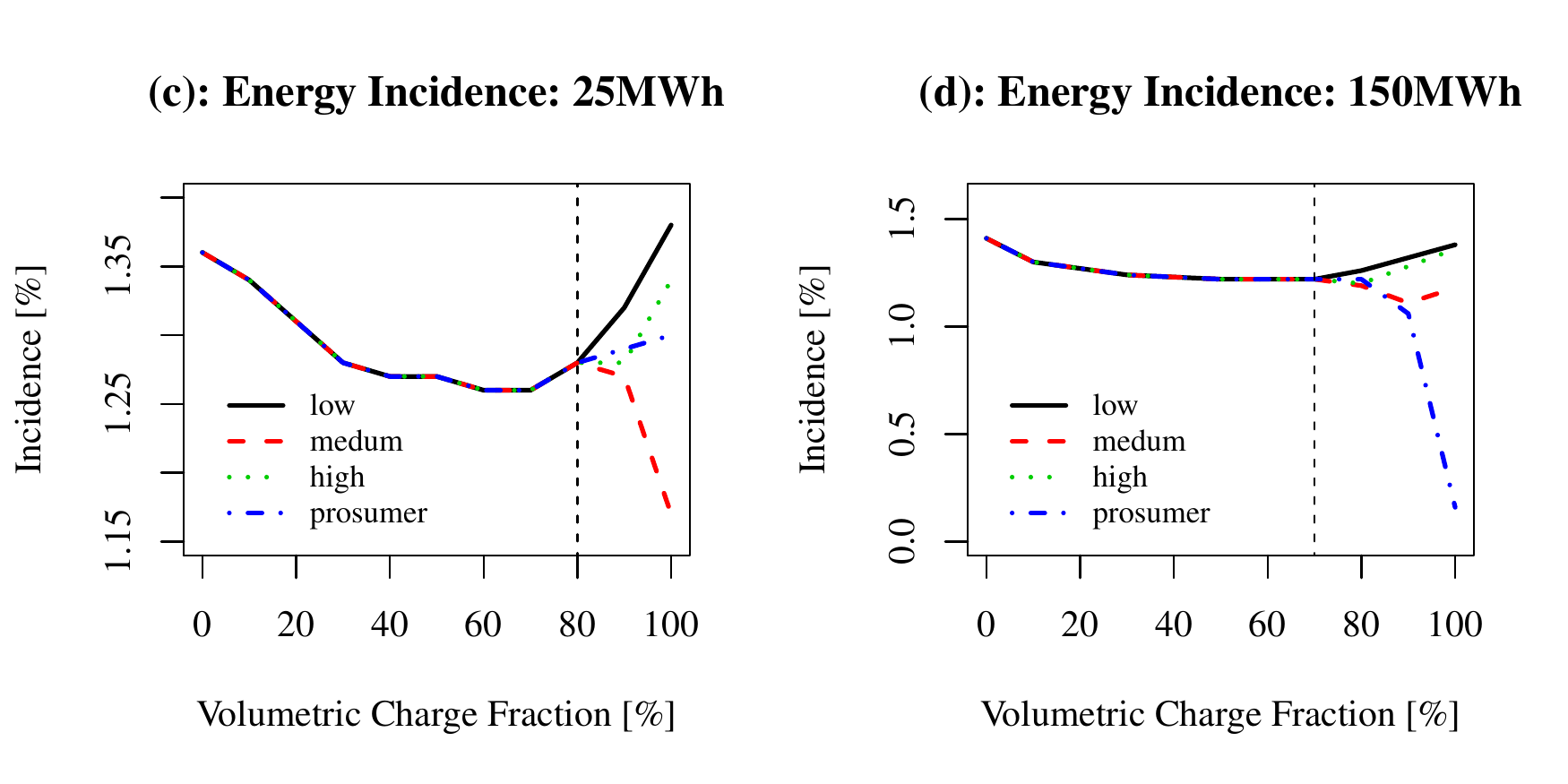}\vspace{-1em}
\caption{Plots of energy incidences}\vspace{-1.5em}
\label{fig:4}
\end{figure}

To zoom in on energy equity, Figure \ref{fig:4} shows the energy incidence by income groups, e.g., low, medium, high and prosumers. When $R = 25$ MWh, 
energy equity can be maintained when the revenue levied from volumetric charge can be less than or equal to $80\% \Tau$; beyond that, the different fixed charges $\phi$ collected from different income groups can no longer be used to maintain energy equity. The similar trend is also seen for $R = 150$ MW.
%We demonstrate that with the allotted volumetric charge fraction less than or equal to 80\% and 70\% in the 25 MWh and 150MWh cases, it is possible to maintain the energy equity.   
%Beyond 80\% and 70\% for these two cases, respectively, the available fund is inadequate to effectively compensate low-income groups, leading to energy inequity. 
\vspace{-1em}

%\begin{figure}[htbp] %  figure placement: here, top, bottom, or page
%\centering
%\includegraphics[width=2in]{./figs/fig2.pdf}
%\caption{Plots of  volumetric tariffs}
%\label{fig:3}
%\end{figure}

\section{Conclusion \vspace{-3pt}}\label{sec:con}
Recovering utilities' fixed costs has presented a significant regulatory challenge in designing electricity tariffs.
While emphasis has typically been placed on economic efficiency and incentive to conserve energy, equally important are their impacts on the energy equity among different income groups.
The situation is further exacerbated by the presence of prosumers who are typically among the most affluent income groups, taking advantage of the electricity tariff, adopting new technologies and optimizing their self-interests.  

This study examines the optimal retail tariff in the presence of prosumers.
We demonstrate that a volumetric approach to recover fixed costs based on energy consumption is likely to be less efficient.
Our analysis concludes that the first-best policy is to leave the wholesale market intact and rely on fixed charge to recover fixed costs. 
However, such a policy prescription is likely implausible:
%subject to regulatory challenge.
lower retail prices will likely provide a disincentive for energy conservation, against the effort to decarbonizing economy.
In addition, a lumpy fixed charge on low-income households can be challenging to those families who already face harsh economic situations. 
Therefore, a policy that provides directly financial compensation to low-income households is economically efficient. 
Our analysis suggests that programs, such as CARE (California Alternate Rate of Energy), designed to mitigate low-income households' energy expenditure via lower retail rates, a volumetric approach, is unlikely to be efficient.

\textcolor{black}{Because our analysis is short-run based, which does not consider the interaction between power-system operations and expansion decisions, 
%e.g, \cite{pozo13}, 
it is subject to a number of long-run implications. 
%More specifically, the second feedback or time-varying rate feedback discussed in \cite{Darghouth16} is not considered in our paper.
%However, if the second effect offsets the benefit of avoiding fixed-cost recovery, we believe that it can further reduce the economic incentive of high-income consumers to convert to prosumers under our proposed fine-tuned policy.
In particular, while the policy is expected to improve conventional consumers' energy expenditure incidence, it may offset their economic incentive to invest in DERs, slowing the development of non-utility-scale DERs.}
\textcolor{black}{Moreover, the impact on incidence can also be affected by demand elasticity.
When demand is less price-responsive, consumers cannot forgo consumption in response to higher power prices, leading to higher consumption and  a lower volumetric tariff.} 
%Finally, the conclusion herein can be applied to the context of other emerging technologies, such as electric vehicle recharging stations.  

{\appendix[Proof of Proposition \ref{pr:OptimalTau}]}
As discussed earlier, $V(\tau^b, \tau^s)$ is the optimal value function of a convex quadratic program with parameters only in the objective function. It is well known that in this case, $V(\tau^b, \tau^s)$ is a convex function (such as Theorem 47 in \cite{TerlakyOptValueFunc}). By the fact regarding maximizing a convex function over a bounded polyhedral set (see Theorem 3.4.7 in \cite{BSS_NLP}), at least one of the extreme points of $\mathcal{T}$ must be an optimal solution of $\max_{(\tau^b, \tau^s) \in \mathcal{T}} V(\tau^b, \tau^s)$. The set $\mathcal{T}$, as defined in \eqref{eq:setT}, has four extreme points: $\tau^I := (\tau^b =0, \tau^s = 0)$, $\tau^{II}:= (\tau^b = \widehat{\tau}^b, \tau^s = \widehat{\tau}^b)$, $\tau^{III}:=(\tau^b = \widehat{\tau}^b, \tau^s = \underline{\tau}^s)$, and 
$\tau^{IV}:= (\tau^b = 0, \tau^s = \underline{\tau}^s)$. In the following, we use $\bm{x}(\tau^e)$ (or an element in the vector of $\bm{x})$, $e = I, II, III$ or $IV$ to denote the optimal solution corresponding to one of the four extreme points. Note that since the feasible region $\bm{X}$ is not perturbed by $\tau$, any optimal solution $\bm{x}(\tau^e)$  will also be feasible for problem $\max_{\bm{x}\in \bm{X}} \Pi(\bm{x}; \tau^{e'})$. 

First, we compare $\Pi(\bm{x}(\tau^I); \tau^I)$ and 
$\Pi(\bm{x}(\tau^{IV}); \tau^{IV})$. Note that since $\underline{\tau}^s$ is defined to be a sufficiently negative number such that $z^s_i$ will be 0 in an optimal solution for any $i=1, \ldots, N$, at an optimal solution $\bm{x}(\tau^{IV})$, $z_i^s (\tau^{IV})* \underline{\tau}^s = 0$, for all $i$. 
In addition, since $\tau^b = 0$ in $\tau^{IV}$,
we have that 
$\Pi(\bm{x}(\tau^{IV}); \tau^{IV}) = \Pi(\bm{x}(\tau^{IV}); \tau^{I}) \leq \max_{\bm{x}\in \bm{X}}
\Pi(\bm{x}; \tau^{I}) = \Pi(\bm{x}(\tau^{I}); \tau^{I}).$
consider $\tau^{II}:= (\tau^b = \widehat{\tau}^b, \tau^s = \widehat{\tau}^b)$. 

Next, we compare $\Pi(\bm{x}(\tau^{II}); \tau^{II})$ and $\Pi(\bm{x}(\tau^{III}); \tau^{III})$. Since $\underline{\tau}^s < \widehat{\tau}^b$, it is easy to see that $\Pi(\bm{x}(\tau^{III}); \tau^{III}) \leq 
\Pi(\bm{x}(\tau^{III}); \tau^{II}) \leq \max_{\bm{x}\in \bm{X}} 
\Pi(\bm{x}; \tau^{II})  = \Pi(\bm{x}(\tau^{II}); \tau^{II})$. 

Finally, we compare $\Pi(\bm{x}(\tau^{I}); \tau^{I})$ and $\Pi(\bm{x}(\tau^{II}); \tau^{II})$. Consider the objective function parameterized at $\tau^{II}$: $\Pi(\bm{x};\tau^{II})$. After rearranging, we have 
\begin{align}
  \Pi(\bm{x}; \tau^{II})& :=      \sum_{i}\Bigg[\int_{0}^{d_i}p^{con}_i(m_i)\, dm_i 
  + \int_{0}^{l_i}p^{pro}_i(m_i)\, dm_i
   + \nonumber \\
  &	- C^g_i(g_i) -  \sum_{h \in H_{i}} C_{ih}(g_{ih}) 
  \Bigg]\nonumber + \sum_i\Bigg[\widehat{\tau}^b (z^s_{i} -  z_i^b -  d_i) \Bigg]. \nonumber
\end{align}
Using the constraint \eqref{eq:nb-price} and \eqref{eq:sumY=0}, we can rewrite the last term above as follows:
\begin{align}\label{eq:KeyObs}
& \widehat{\tau}^b\sum_i\Bigg(z^s_{i} -  z_i^b -  d_i\Bigg) = \widehat{\tau}^b\sum_i\Bigg(y_i - \sum_{h\in H_i} g_{ih}\Bigg) \nonumber \\
& =  \widehat{\tau}^b\Bigg[\sum_i y_i - 
\Bigg(\sum_i\sum_{h\in H_i} g_{ih}\Bigg)\Bigg]  = - \widehat{\tau}^b \Bigg(\sum_i\sum_{h\in H_i} g_{ih}\Bigg). \nonumber
\end{align}
Since $\widehat{\tau}^b > 0$ and $g_{ih} \geq 0$ for all $i$ and $h\in H_i$, by replacing $\widehat{\tau}^b$ with 0, we can obtain a no smaller objective function value. As a result, we have 
\begin{equation}
\begin{aligned}
\Pi(\bm{x}( \tau^{II}); \tau^{II})\ 
& \leq\ \Pi(\bm{x}( \tau^{II}); \tau^s = \tau^b = 0) \\
&\leq\ \max_{\bm{x}\in \bm{X}} 
\Pi(\bm{x}; \tau^{I})  = \Pi(\bm{x}(\tau^{I});(\tau^{I})) . 
\end{aligned}
\end{equation}
Hence, we have that $\tau^{I} = (0,0) \in \arg\max_{\tau\in\mathcal{T}}V(\tau)$. 
%By the arguments regarding the linear system equations 
%\eqref{eq:EqualIncidence} in the proof of Proposition \ref{pr:Equiv}, we know that with any given $\tau$, there exists $(\phi^{con}, \phi^{pro})$ to make the energy equity index zero. Let $(\phi^{con}(0,0),  \phi^{pro}(0,0))$ denote one particular solution to \eqref{eq:EqualIncidence} with $\tau^s = \tau^b = 0$. Then $\zeta^* = (0, 0, \phi^{con}(0,0), \phi^{pro}(0,0))$ is an optimal solution to the BLP 
%\eqref{eq:UpperObj} -- \eqref{eq:UpperLastConstr}. 
\hfill $\Box$ \vspace{1em}

%\section*{Acknowledgment}


\bibliographystyle{IEEEtran}
\bibliography{optimal}

\end{document}